\documentstyle[aps,prc,eqsecnum,epsfig,floats]{revtex}
\tightenlines
\newcommand \ie {{\it i.e.}}
\newcommand \f {\not\!}
\newcommand \wh {\widehat}

\newcommand \hk {\hat{k}}
\newcommand \hw {\hat{w}}
\newcommand \nf {\tilde{n}} 
\newcommand \kq {E_{k-q}}
\newcommand \qk {E_{q-k}}
\newcommand \uqk {\widehat{(q-k)}}

\newcommand \kd  {\delta}
\newcommand \ra  {\rightarrow}
\newcommand \ev {\delta \Gamma}
\newcommand \po {$\Pi^{1}$}
\newcommand \pt {$\Pi^{2}$}
\newcommand \w  {\omega}
\newcommand \fw {{\bf w}}
\newcommand \fp {{\bf p}}
\newcommand \fk {{\bf k}}
\newcommand \fq {{\bf q}}
\newcommand \h {\theta}
\newcommand \im {\Rightarrow}
\newcommand \vk {\vec{k}}
\newcommand \vq {\vec{q}}
\newcommand \vw {\vec{w}}
\newcommand \lra {\leftrightarrow}
\newcommand \mat {{\mathcal M}}
\newcommand \g {\gamma}
\newcommand \r {\rho}

\newcommand \e {\epsilon}
\newcommand \onp {(1/2 - \tilde{n}(k))^{\prime}}
\newcommand \p {^{\prime}}
\newcommand \N {{\mathcal N}}
\newcommand \Sc {{\mathcal S}}
\newcommand \x {\cdot}
\newcommand \hf {\frac{1}{2}}
\newcommand \A {\alpha}
\newcommand \B {\beta}
\newcommand \tbe { \log \left( \frac{2}{\e} \right) }

\begin{document}

\draft

\title{ On the imaginary parts and infrared divergences of 
two-loop vector boson self-energies in thermal QCD }

\author{A. Majumder and C. Gale}

\address{
Physics Department, McGill University, 3600 University St.,\\ Montr{\'e}al, QC.
Canada H3A 2T8}

\date{ \today}

\maketitle

\begin{abstract}
We calculate the imaginary part of the retarded two-loop 
self-energy of a static vector
boson in a plasma of quarks and gluons of temperature $T$, using  
the imaginary time formalism.
We recombine the various cuts of the
self-energy to generate physical processes. We demonstrate how 
cuts containing loops may be reinterpreted in terms of interference between 
$O(\A)$ tree diagrams and the Born term along with spectators from the 
medium. 
We apply our results to 
the rate of dilepton production in the limit of dilepton invariant
mass $E>>T$. 
We find that all infrared and collinear singularities 
cancel in the final result obtained in this limit.
\end{abstract}


\section{Introduction}

The imaginary parts of retarded self-energies represent 
extremely important quantities in thermal field theory. They 
provide information about various quantities of physical 
interest in the medium. Primary among these are the 
decay and formation rates of particles \cite{wel83}. 
Boson self-energies provide information about quantities 
like $Z$ decay rates \cite{kap00b}, and production rates of
dileptons and real photons \cite{toimc} from a 
quark-gluon-plasma (QGP).  
The spectrum of lepton pairs ({\it i.e.}, $e^+e^-$, $\mu^+\mu^-$)
 and real photons 
emanating from such a plasma has been considered as a promising signature of
QGP formation \cite{shu80,kaj86}. This owes to the fact that the 
photons or dileptons suffer essentially no
final state interaction.

Some years ago the 
contribution to the rate of dileptons produced at rest in the plasma 
at first order in 
the strong coupling constant, was evaluated \cite{bai88}. 
This included reactions like three particle
fusion ($q\bar{q}g \ra \g^*$), Compton scattering ($ q g \ra q \g^*$ or $ \bar{q}
g \ra \bar{q} \g^*$), pair annihilation ($q\bar{q} \ra g \g^*$), Born term with
vertex correction, and Born term with quark or antiquark 
self-energy correction. 
This calculation was performed in the real time 
formalism, both in a Feynman
diagram approach in thermo-field dynamics, and by 
taking the imaginary part of
the two loop photon self-energy. In the case of massless QCD, 
each of the contributions mentioned above contain
infrared or collinear singularities. These were 
regulated at intermediate stages of
the calculation by giving masses to the quarks 
and gluons. The combined rate from
all these processes was then found to be free 
of all divergences in the limit of
vanishing masses. This calculation was also performed 
simultaneously by another group
 \cite{alt89}, who dimensionally regularized the
singularities at intermediate stages of the calculation. 
The end result remained the
same: when all the different processes were summed, 
the divergences cancelled and
dilepton rate at next-to-leading order remained finite.

Recent calculations employing a multiple scattering
expansion, however, have found a remnant collinear 
divergence \cite{kap00b,kap00}. 
This result has been commented upon \cite{aur02}, and the issue of 
divergences remained unresolved \cite{kap02}. 
In the wake of this strife, we revisit this problem in a systematic calculation.
Also, to the best of our knowledge, 
a complete calculation of the imaginary part of a heavy 
vector boson
retarded self-energy in the imaginary time formalism 
has yet to be performed. This is the subject of this paper. 
The scalar boson self-energy was examined recently \cite{kap01}. 
There are various advantages
to such a calculation: the basic Feynman rules are easily generalized from
zero-temperature; there is no doubling of degrees of freedom and no 
matrix structure of
propagators; multiple poles which lead to ill-defined products of delta
functions in the real-time formalism are easily and 
naturally handled both in the
Matsubara sums and in the analytic continuation. The purpose of this
calculation is thus many-fold. A first goal is 
to enumerate and interpret the various 
physical contributions
contained in the imaginary part of the two-loop self-energies. 	In
doing this, it shall then be shown that  
cuts containing loops may be re-expressed as interference between 
tree diagrams and Born term with a thermal medium spectator. 
Importantly, we also demonstrate how double 
poles may be simply and elegantly dealt with,
in the Matsubara sum and in the analytic continuation to real energies. 
We finally concentrate on eventual 
collinear and infrared divergences in the ensuing rates. In this
study, we focus on the singularity structure in the region of phase space 
investigated by the authors of Refs. \cite{kap00b,kap00}. 
Even though we explicitly calculate the self-energies of static virtual 
photons, the results may be easily applied to other vector bosons in-medium,
with the exception of the gluon which admits other self-energies 
in a QGP. 

The various sections 
are organized as follows: in section II we begin by 
evaluating one of the self-energy
diagrams of a static photon with an 
imaginary energy at two loops (the impatient reader may
skip ahead to section VI where the various cuts 
of the self-energy are recombined 
to provide physical interpretations of the various terms 
obtained; following which the 
infra-red behaviour of 
heavy photon production will be discussed); 
in section III we analytically continue the photon energy 
to real values and
obtain the imaginary part of the corresponding retarded 
self-energy; in section IV 
we evaluate the other self-energy topology; in section V we analytically 
continue this 
self-energy to real values of photon energy and find the 
retarded imaginary self-energy; 
in section VI we combine the tree-like cuts 
and reinterpret
them as physical processes with thermal distributions 
on the phase space factors; in section
VII we attempt to interpret cuts containing loops in 
terms of the recently proposed spectator
interpretation \cite{won01}; in section VIII we take 
the limit of heavy-photon production ($E>>T$), 
and evaluate the various contributions; in section IX we
combine all cuts, demonstrate the cancellation of the collinear and 
infrared divergences, and
present our results; we present our conclusions and 
brief discussions in section X.
A few short appendices follow. In the interest of quantitative
accuracy and repeatability, we have presented many 
calculational details: the issue being addressed 
here is technical, and thus demands a rigorous treatment.

\section{The self-energy: Topology I}

\begin{figure}[htbp]
  \begin{center}
  \epsfxsize 80mm
  \epsfbox{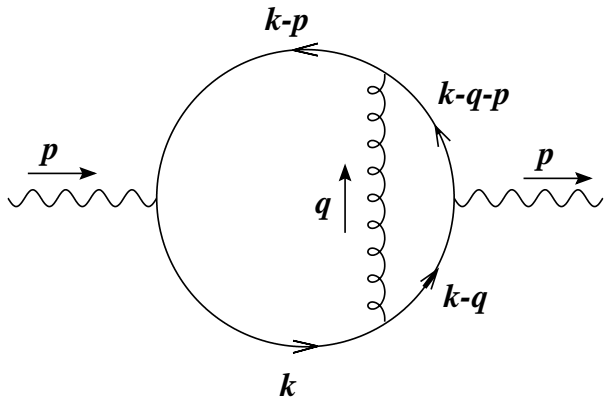}
\vspace{0.25cm}
    \caption{The first topology for the self-energy.}
    \label{se}
  \end{center}
\end{figure}

We evaluate the photon self energy with a gluon running across as shown in
Fig.~\ref{se}. To begin with, we derive the expression for the effective
quark photon vertex corrected by a gluon running across, \ie,

\[
ie\Gamma^\mu = ie\gamma^\mu + ie\kd \Gamma^\mu.
\]

Where $e$ may be taken to be the electric charge of the quark.
In standard notation, the expression for the effective vertex 
in Feynman gauge may be written down as

\begin{equation}
ie\Gamma^{\mu} = \frac{i}{\beta} \sum_{q^0} \int \frac{d^3 q}{(2\pi)^3}
\frac{-ig_{\rho \sigma} \kd^{a b}}{q^2} 
(it_{i,j}^{a}g\gamma^{\rho})
\frac{i(\f k-\f q-\f p)}{( k-q-p )^2}
(ie\gamma^{\mu})
\frac{i(\f k-\f q)}{(k-q)^2}
(it_{j,k}^{b} g\gamma^{\sigma}).
\end{equation}

The Matsubara sum in the effective vertex  
may be simply evaluated using the method of Pisarski \cite{pis88}. 
In our notation (see Appendix A, see also \cite{maj01}), 
this is given in the static limit 
$(\vec{p}=0)$ as,

\begin{eqnarray}
\ev^\mu &=& \frac{ g^2 C_{ik}}{4} \int \frac{d^3 q}{(2\pi)^3} \sum_{s_1,s_2,s_3}
\frac{( \f k - \f q )_{s_2} \gamma^{\mu} ( \f k - \f q )_{s_3} }
{ q \qk \qk  ( p^0 - (s_2-s_3)E_{q-k} )} \nonumber \\
&\times & \Bigg\{ 
-s_3 \frac{(s_1+s_2)/2 - s_1 \nf(\qk) + s_2 n(q) }{k^0 - s_1q - s_2\qk} + 
s_2 \frac{(s_1+s_3)/2 - s_1 \nf(\qk) + s_3 n(q) }{k^0-p^0 - s_1q - s_3\qk} 
\Bigg\}. \label{effvert}
\end{eqnarray}

Where $s_1,s_2,s_3$ are sign factors which are summed over the values of
$\pm 1$. 
We may now use the above result to write the full 
self-energy of the
photon in the static limit as 

\begin{eqnarray}
i\Pi_{\mu}^{\mu} &=& \frac{i}{\beta} \sum_{k^{0}} \int  
\frac{ d^3 k}{ (2\pi)^{3} }
(-1)Tr \sum_{s s_4} \Bigg[ e \gamma_{\mu} \kd_{ki}
\frac{ \gamma^\beta s_4 \hk_{ \beta , s_4 } }{ 2(k^0-p^0 - s_4 k) } 
e \ev^{\mu}_{i,k}
\frac { \gamma^\alpha s \hk_{ \alpha , s } }{ 2(k^0 - s k) }
\Bigg].
\end{eqnarray}

Where, $\hk_{s}$ stands for the four component quantity:

\[
\Bigg\{ s, \frac{k_x}{k}, \frac{k_y}{k}, \frac{k_z}{k} \Bigg\} = \Big\{ s,0,0,1
\Big\}.
\]

We note that the effective
vertex may be written as

\[
\ev^{\mu} = \gamma^{\rho} \gamma^{\mu} \gamma^{\sigma} \ev_{\rho \sigma},
\]

to highlight the structure of $\gamma$ matrices 
contained within it. The trace of the 
$\gamma$ matrices is straightforward.
This gives the full self energy as

\begin{eqnarray}
\Pi_{\mu}^{\mu} &=& \frac{-4}{\beta} \sum_{k^{0}} \int  
\frac{ d^3 k}{ (2\pi)^{3} }
\sum_{s s_4} e^2 \Bigg[   
\frac { s \hk_{ \alpha , s } }{ k^0 - s k }
 \ev^{\beta \alpha}
\frac{ s_4 \hk_{ \beta , s_4 } }{ k^0-p^0 - s_4 k } 
\Bigg]. \label{slfeng1sht}
\end{eqnarray}

For convenience we change $ s_2 \ra -s_3 $ and $ s_3 \ra -s_2 $. 
To evaluate the Matsubara sum we follow the method of reference 
\cite{kap89}. This method converts the Matsubara sum into a contour 
integration in the
complex plain of $k^0$, \ie,

\begin{eqnarray}
\Pi_{\mu}^{\mu} &=& \frac{-4 e^2 g^2}{2 \pi i}  
\int_{-i\infty + \epsilon}^{i\infty + \epsilon} dk^0
\int \frac{ d^3 k d^3 q}{ (2\pi)^{6} }
[ 1/2 - \nf(k^0) ]\frac { s \hk_{ \alpha , s } }{ s_5k^0 - s k } \nonumber \\ 
&\times & \Bigg[
\frac{ \uqk_{s_2}^{\alpha} \uqk_{s_3}^{\beta} }
{  q ( p^0 - (s_2-s_3)E_{q-k} )} 
\Bigg\{ 
s_2 \frac{(s_1-s_3)/2 - s_1 \nf(\qk) - s_3 n(q) }{s_5k^0 - s_1q + s_3\qk} 
\nonumber \\ 
&+& s_3 \frac{(s_1-s_2)/2 - s_1 \nf(\qk) - s_2 n(q) }
{ p^0 - s_5k^0 + s_1q - s_2\qk} 
\Bigg\} \Bigg]
\frac{ s_4 \hk_{ \beta , s_4 } }{ s_5k^0 - p^0 - s_4k }. \label{slfeng1lng}
\end{eqnarray}

Where, $\uqk_{s}$ stands for the four component quantity:
\[
\Bigg\{ s, \frac{q_x - k_x}{|q-k|}, \frac{q_y - k_y}{|q-k|}, \frac{q_z - k_z}{|q-k|}
\Bigg\} 
= \Bigg\{ s, \frac{q\sin\h \cos \phi}{\sqrt{k^2 + q^2 - 2kq\cos\h}}, 
\frac{q\sin\h \sin \phi}{\sqrt{k^2 + q^2 - 2kq\cos\h}},
\frac{q\cos\h-k}{\sqrt{k^2 + q^2 - 2kq\cos\h}}
\Bigg\}.
\]

The $k^0$ integration is from $-\infty \rightarrow \infty$ on the positive 
side of the real axis. We may thus close the contour on the positive side. 
Note that the function is vanishing as $k^0 \rightarrow \infty$. 
The result of this integration will simply be the sum of the residues at 
the corresponding poles. Looking at the above expression we note that the pole
structure is different depending on whether the term being considered is the
first or the second one in the curly brackets. We note the following poles:

\vspace{0.5cm}

i) 1st order pole: at $k^0 = k$, requires $s_5 = s$ (in both terms). 

\vspace{0.5cm}

ii) 1st order pole: at $k^0 = k + s_5 p^0$, requires $s_5 = s_4$ (in both
terms).

\vspace{0.5cm}

iii) 1st order pole: at $k^0 = s_5s_1 q - s_5s_3 \qk$, requires 
$s_5s_1 q - s_5s_3 \qk > 0 $ (only in the first term).

\vspace{0.5cm}

iv) 1st order pole: at $k^0 = s_5 p^0 - s_5 s_2 \qk + s_5s_1 q$ (only in the
second term).

\vspace{0.5cm}

In the following, each of the poles are evaluated in a separate subsection and
then summed up. 


\subsection{1st order pole at $k^0 = k$}

This is the pole of the first outer propagator (\ie, not a propagator 
in the effective vertex), it is a pole for the entire self energy 
expression. It has the obvious residue of

\begin{eqnarray}
\Pi_{\mu}^{\mu}(A) &=&  4 e^2 g^2  
\int \frac{ d^3 k d^3 q}{ (2\pi)^{6} }
[ 1/2 - \nf(k) ]\frac { s \hk_{ \alpha , s } }{ s} \nonumber \\ 
&\times & \Bigg[
\frac{ \uqk_{s_2}^{\alpha} \uqk_{s_3}^{\beta} }
{  q ( p^0 - (s_2-s_3)E_{q-k} )} 
\Bigg\{ 
s_2 \frac{(s_1-s_3)/2 - s_1 \nf(\qk) - s_3 n(q) }{sk - s_1q + s_3\qk} 
\nonumber \\ 
&+& s_3 \frac{(s_1-s_2)/2 - s_1 \nf(\qk) - s_2 n(q) }
{ p^0 - sk + s_1q - s_2\qk} 
\Bigg\} \Bigg]
\frac{ s_4 \hk_{ \beta , s_4 } }{ sk - p^0 - s_4k }. 
\end{eqnarray}
Note that there is an extra negative sign in the residue as the contour is 
being taken in the clockwise sense. 

\subsection{1st order pole at $k^0 = k + s_5p^0$}

This is the pole of the second outer propagator, it is a 
pole for the entire self energy expression. 
It gives the residue,

\begin{eqnarray}
\Pi_{\mu}^{\mu}(B) &=&  4 e^2 g^2  
\int \frac{ d^3 k d^3 q}{ (2\pi)^{6} }
[ 1/2 - \nf(k) ]\frac { s \hk_{ \alpha , s } }{ p^0 + s_4 k- s k } \nonumber \\ 
&\times & \Bigg[
\frac{ \uqk_{s_2}^{\alpha} \uqk_{s_3}^{\beta} }
{  q ( p^0 - (s_2-s_3)E_{q-k} )} 
\Bigg\{ 
s_2 \frac{(s_1-s_3)/2 - s_1 \nf(\qk) - s_3 n(q) }{p^0 + s_4 k - s_1q + s_3\qk} 
\nonumber \\ 
&+& s_3 \frac{(s_1-s_2)/2 - s_1 \nf(\qk) - s_2 n(q) }
{ -s_4k + s_1q - s_2\qk} 
\Bigg\} \Bigg]
\frac{ s_4 \hk_{ \beta , s_4 } }{ s_4 }. 
\end{eqnarray}

Note that the $p^0$ in the distribution function has been dropped. 
This may be
done as $e^{p^0 \beta} = 1$ ($p^0$ is a discrete even frequency), 
and, secondly, as we are eventually going to analytically continue the 
self-energy to complex values of $p^0$. The correct analytic continuation is
given by that function which has no non-analytic behaviour off the real axis
\cite{bay61}. 
One may easily check that the above function with a $p^0$ in the distribution
function will have poles at $p^0 = -k + i 2 (n+1) \pi T$.
Note that in this pole we may switch 
\[
s \ra -s_4,\mbox{ } s_4 \ra -s,\mbox{ } s_2 \ra -s_3,\mbox{ } s_3 \ra -s_2,
\mbox{ } s_1 \ra -s_1,
\]
and noting that $\hk_{-s}\uqk_{-s_2}=\hk_{s}\uqk_{s_2}$, we find 
\[
 \Pi_{\mu}^{\mu}(B) = \Pi_{\mu}^{\mu}(A).  
\]

\subsection{1st order pole at $k^0 = s_5s_1 q - s_5s_3 \qk$ } 

In the expression for the effective vertex (Eq.~(\ref{effvert}) 
or Eq.~(\ref{slfeng1lng})), 
we note the presence of two terms inside the curly brackets 
with different pole structures. 
This is a pole of the first term, and is realized only if 
$s_5s_1 q - s_5s_3 \qk > 0 $. 
Thus the residue is 

\begin{eqnarray}
\Pi_{\mu}^{\mu}(C) &=& 4 e^2 g^2  
\int \frac{ d^3 k d^3 q}{ (2\pi)^{6} }
[ 1/2 - \nf(s_5s_1 q - s_5s_3 \qk) ]
\frac { s \hk_{ \alpha , s } }{ s_1 q - s_3 \qk  - s k } \nonumber \\ 
& & \!\!\!\!\!\!\!\!\!\!\!\!\times \Bigg[
\frac{ \uqk_{s_2}^{\alpha} \uqk_{s_3}^{\beta} }
{  q ( p^0 - (s_2-s_3)E_{q-k} )} 
\Bigg\{ 
s_2 \frac{(s_1-s_3)/2 - s_1 \nf(\qk) - s_3 n(q) }{s_5} 
\Bigg\} \Bigg]
\frac{ s_4 \hk_{ \beta , s_4 } \Theta(s_5s_1 q - s_5s_3 \qk)}{ s_1 q - s_3 \qk - p^0 - s_4k }. 
\end{eqnarray}

\subsection{1st order pole at $ k^0 = s_5p^0 + s_5s_1 q - s_5s_2 \qk $ } 

This is the pole of the second term in the curly bracket mentioned in the previous section. 
It is realized only if $s_5s_1 q - s_5s_2 \qk > 0 $. The residue is

\begin{eqnarray}
\Pi_{\mu}^{\mu}(D) &=& 4 e^2 g^2 
\int \frac{ d^3 k d^3 q}{ (2\pi)^{6} }
[ 1/2 - \nf(s_5s_1 q - s_5s_2 \qk) ]
\frac { s \hk_{ \alpha , s } }{ p^0 + s_1 q - s_2 \qk - s k } \nonumber \\ 
& & \!\!\!\!\!\!\!\!\!\!\!\!\times \Bigg[
\frac{ \uqk_{s_2}^{\alpha} \uqk_{s_3}^{\beta} }
{  q ( p^0 - (s_2-s_3)E_{q-k} )} 
\Bigg\{ 
 s_3 \frac{(s_1-s_2)/2 - s_1 \nf(\qk) - s_2 n(q) }
{-s_5} 
\Bigg\} \Bigg]
\frac{ s_4 \hk_{ \beta , s_4 } \Theta(s_5s_1 q - s_5s_2 \qk)}
{s_1 q - s_2 \qk  - s_4k }. 
\end{eqnarray}

 Note that in this pole we may switch 

\[
s_2 \ra -s_3, \mbox{ } s_3 \ra -s_2, \mbox{ } s_1 \ra -s_1,\mbox{ }  
s_5 \ra -s_5,\mbox{ } s \ra -s_4,\mbox{ } s_4 \ra -s. 
\]

With this operation, we find 

\[
 \Pi_{\mu}^{\mu}(D) = \Pi_{\mu}^{\mu}(C).  
\]

Thus the full photon self-energy to second order 
in the coupling constant for the diagram of 
Fig.~\ref{se} is given by summing up the results of the
preceding four subsections, \ie,

\[
\Pi_{\mu}^{\mu} = 2\Pi_{\mu}^{\mu}(A) + 2\Pi_{\mu}^{\mu}(C). 
\]


\section{Imaginary part of the first self-energy topology}


We now proceed with evaluating the discontinuity in the first self-energy 
as $p^0$ is analytically continued towards the positive real axis from above 
\ie, 
$p^0 \ra E + i\epsilon$. 
Analytically continuing $p^0$ will give us the retarded 
self energy of the photon in real time in terms of a real 
continuous energy $p^0=E$. 
The expressions to be continued are $\Pi_{\mu}^{\mu}(A)$ and $\Pi_{\mu}^{\mu}(D)$.
The presemce of the theta function in $\Pi_{\mu}^{\mu}(D)$ complicates the pole 
sturcture that one would obtain during analytic continuation of this expression. 
Using standard techniques (outlined in Sect. IV. D.), 
we decompose the theta function to obtain,

\begin{eqnarray}
\Pi_{\mu}^{\mu}(D) &=& -4 e^2 g^2 
\int \frac{ d^3 k d^3 q}{ (2\pi)^{6} }
\frac{ss_4s_3}{q} \nonumber \\
& & \!\!\!\!\!\!\!\!\!\!\!\!\!\!\!\!\!\!\!\!\!\!\!\!\!\!\!\!\!\!\!\!\!\!\!\!\!\!\!\!
\times \Bigg\{ \frac{[\hk_s \cdot \uqk_{-s_5}][ \hk_{s_4} \cdot \uqk_{s_3}]
[ 1 - \nf(\qk) + n(q) ][ 1/2 - \nf( \qk + q ) ]}{[p^0 - (sk - s_5\qk - s_5q)] 
[p^0 - ( -s_5 - s_3 )\qk ][ s_5( \qk + q ) - s_4 k ]} \nonumber \\
& & 
\!\!\!\!\!\!\!\!\!\!\!\!\!\!\!\!\!\!\!\!\!\!\!\!\!\!\!\!\!\!\!\!\!\!\!\!\!\!\!\!
\mbox{} - \frac{[\hk_s \cdot \uqk_{s_5}][ \hk_{s_4} \cdot \uqk_{s_3}]
[ \nf(\qk) + n(q) ][ 1/2 - \nf( q - \qk ) ]}{[p^0 - (sk + s_5\qk - s_5q)]
[p^0 - ( s_5 - s_3 )\qk ][ s_5( q - \qk ) - s_4 k ] }
\Bigg\}.
\end{eqnarray}

Analyzing the expressions for $\Pi_{\mu}^{\mu}(A)$ and $\Pi_{\mu}^{\mu}(D)$, 
we note the following discontinuities:

\vspace{0.5cm}

Poles of type $p^0 = 2k$:

\vspace{0.5cm}

i) 1st order pole in $\Pi_{\mu}^{\mu}(A)$: at $p^0=2k$, 
requires $s=-s_4=1$ (in both terms that make up $\Pi_{\mu}^{\mu}(A)$).

\vspace{0.5cm}

Poles of type $p^0 = 2E_{q-k}$:

\vspace{0.5cm}

ii) 1st order pole in $\Pi_{\mu}^{\mu}(A)$: at $p^0=2E_{q-k}$, 
requires $s_2=-s_3=1$ (in both terms that make up $\Pi_{\mu}^{\mu}(A)$).

\vspace{0.5cm}

iii) 1st order pole in $\Pi_{\mu}^{\mu}(D)$: at $p^0=2E_{q-k}$, 
requires $s_5=s_3=-1$ (only in first term of $\Pi_{\mu}^{\mu}(D)$).

\vspace{0.5cm}

iv) 1st order pole in $\Pi_{\mu}^{\mu}(D)$: at $p^0=2E_{q-k}$, 
requires $s_5=-s_3=1$ (only in second term of $\Pi_{\mu}^{\mu}(D)$). 

\vspace{0.5cm}

Poles of type $p^0 = sk + s_1q + s_2\kq$:

\vspace{0.5cm}

v)  1st order pole in $\Pi_{\mu}^{\mu}(A)$: at $p^0= sk - s_1q + s_2\qk$, 
requires $sk - s_1q + s_2\qk>0$ (only in the second term of $\Pi_{\mu}^{\mu}(A)$).   

\vspace{0.5cm}

vi) 1st order pole in $\Pi_{\mu}^{\mu}(B)$: at $p^0= sk - s_5q - s_5\qk$, 
requires $sk - s_5q - s_5\qk>0$ (only in the first term of $\Pi_{\mu}^{\mu}(B)$).

\vspace{0.5cm}

vii) 1st order pole in $\Pi_{\mu}^{\mu}(B)$: at $p^0= sk - s_5q + s_5\qk$, 
requires $sk - s_5q + s_5\qk>0$ (only in the second term of $\Pi_{\mu}^{\mu}(B)$).

\vspace{0.5cm}

We may write down the expression for $2\Pi_{\mu}^{\mu}(A)$ 
highlighting its real and imaginary parts as $p^0 \ra E + i\e$ as 

\begin{eqnarray}
2\Pi_{\mu}^{\mu}(A)] &=& -8e^2g^2 
\int \frac{ d^3 k d^3 q }{ (2\pi)^6 }
\frac{ s_4 [\hk_s \cdot \uqk_{s_2}][ \hk_{s_4} \cdot \uqk_{s_3} ] }  
{  E_{q+k} } \nonumber \\
&\times & \Bigg[ 
\mbox{P} \left( \frac{1}{ E - (s_2-s_3)E_{q-k} } \right) - i\pi \kd(E- ( s_2 - s_3 )E_{q-k}) \Bigg]
\nonumber \\
&\times & \Bigg\{ 
s_2 \frac{(s_1-s_3)/2 - s_1 \nf(\qk) - s_3 n(q) }{sk - s_1q + s_3\qk}  
+ s_3 ((s_1-s_2)/2 - s_1 \nf(\qk) - s_2 n(q) ) \nonumber \\
&\times & \Bigg[ 
\mbox{P} \left( \frac{ 1 }{ E - sk + s_1q - s_2\qk } \right) - i\pi \kd ( E - sk + s_1q - s_2\qk ) 
\Bigg]
\Bigg\}
[ 1/2 - \nf(k) ] \nonumber \\
&\times & \Bigg[
\mbox{P} \left( \frac{ 1 }{  E - ( s - s_4 )k } \right) - i\pi \kd ( E - ( s - s_4 )k )
\Bigg]. 
\end{eqnarray}

We now write down the various discontinuities as enumerated above 
(Note: we are now looking at $2\Pi_{\mu}^{\mu}(A)$ and $2\Pi_{\mu}^{\mu}(D)$, 
so the overall factors have doubled):

\vspace{1cm}

\begin{eqnarray}
Disc[2\Pi_{\mu}^{\mu}(A)]_a &=&  (+2 \pi i) 8 e^2 g^2  
\int \frac{ d^3 k d^3 q}{ (2\pi)^{6} }
\frac{ -2 [\hk_+ \cdot \uqk_{s_2}][ \hk_{-} \cdot \uqk_{s_3} ] }
{  q ( E - (s_2-s_3)E_{q-k} )}  \nonumber \\
& & \!\!\!\!\!\!\!\!\!\! \times
s_2 \frac{(s_1-s_3)/2 - s_1 \nf(\qk) - s_3 n(q) }{k - s_1q + s_3\qk}  
[ 1/2 - \nf(k) ] \kd(E-2k). \label{discAa}
\end{eqnarray}

\begin{eqnarray}
Disc[2\Pi_{\mu}^{\mu}(A)]_b &=&  (+2 \pi i) 8 e^2 g^2  
\int \frac{ d^3 k d^3 q}{ (2\pi)^{6} }
\frac{ s_4 [\hk_s \cdot \uqk_{+}][ \hk_{s_4} \cdot \uqk_{-} ] }
{ q }  \nonumber \\
&\times & \Bigg\{ 
\frac{[ 1 - 2\nf(\qk) ]2q}{( sk - \qk )^2 - q^2 }
\Bigg\} 
\frac{ [ 1/2 - \nf(k) ] }{ E - ( s - s_4 )k } \kd(E-2\kq).  \label{discAb}
\end{eqnarray}

\begin{eqnarray}
Disc[2\Pi_{\mu}^{\mu}(D)]_b &=& (+2 \pi i) 8 e^2 g^2 
\int \frac{ d^3 k d^3 q}{ (2\pi)^{6} }\,
\frac{1}{q}\,
\Bigg\{ 
\frac{[\hk_s \cdot \uqk_{+}][ \hk_{s} \cdot \uqk_{-}]}{\qk^2 - (sk+q)^2}
\nonumber \\
&+& \frac{[\hk_s \cdot \uqk_{+}][ \hk_{s} \cdot \uqk_{+}]}{q^2-(\qk-sk)^2}
\Bigg\} 
\Big\{ 
[1-\nf(\qk) + n(q)][1/2 - \nf(\qk + q)] \nonumber \\
&-& [\nf(\qk) + n(q)][1/2 - \nf(q - \qk)] \Big\} \kd(E-2\qk). \label{discDa}
\end{eqnarray}

We now proceed to the evaluations of the discontinuities of the 
type $p^0 = sk + s_2\qk + s_1q$. 
We first change $s_1 \ra -s_1$ in 
$2\Pi_{\mu}^{\mu}(A)$, and $s_5 \ra -s_5$ in the first part of 
$2\Pi_{\mu}^{\mu}(D)$. Hence the discontinuity occurs in 
$2\Pi_{\mu}^{\mu}(A)$ at $p^0 = sk + s_2\qk + s_1q$ only when 
$sk + s_2\qk + s_1q>0$. This may happen in only one of four instances: 
$s=+,s_2=+,s_1=+$; $s=-,s_2=+,s_1=+$, $s=+,s_2=-,s_1=+$; $s=+,s_2=+,s_1=-$. 
In $2\Pi_{\mu}^{\mu}(D)$ the discontinuity occurs in the 
first term when $s=+,s_5=+$ or when $s=-,s_5=+$; in 
the second term when $s=+,s_5=+$ or when $s=+,s_5=-$.

Thus the discontinuity in $2\Pi_{\mu}^{\mu}(A)$ occurs in four parts:

\begin{eqnarray} 
Disc[2\Pi_{\mu}^{\mu}(A)]_c &=& (+2\pi i) 8 e^2 g^2  
\int \frac{ d^3 k d^3 q}{ (2\pi)^{6} q}  [1/2 - \nf(k)]  \label{discAc} \\
\times &\Bigg\{& \frac{s_3 s_4 [\hk_{+} \cdot \uqk_+] [\hk_{s_4} \cdot \uqk_{s_3}]
[-1 + \nf(\qk) - n(q)]}{[ k + q + s_3\qk ][ q + \qk + s_4k] } \kd(E-k-\qk-q)
\nonumber \\
&+& \frac{s_3 s_4 [\hk_{-} \cdot \uqk_+] [\hk_{s_4} \cdot \uqk_{s_3}]
[-1 + \nf(\qk) - n(q)]}{[ -k + q + s_3\qk ][ q + \qk + s_4k] } \kd(E+k-\qk-q)
\nonumber \\
&+& \frac{s_3 s_4 [\hk_{+} \cdot \uqk_-] [\hk_{s_4} \cdot \uqk_{s_3}]
[\nf(\qk) + n(q)]}{[ k + q + s_3\qk ][ -\qk + q + s_4k] } \kd(E-k+\qk-q)
\nonumber \\
&+&\frac{s_3 s_4 [\hk_{+} \cdot \uqk_+] [\hk_{s_4} \cdot \uqk_{s_3}]
[-\nf(\qk) - n(q)]}{[ k - q + s_3\qk ][ \qk - q  + s_4k] } \kd(E-k-\qk+q)
\Bigg\}. \nonumber 
\end{eqnarray} 

The discontinuity in $2\Pi_{\mu}^{\mu}(D)$ also occurs in four parts: 

\begin{eqnarray} 
Disc[2\Pi_{\mu}^{\mu}(D)]_c &=& (-2\pi i) 8 e^2 g^2  
\int \frac{ d^3 k d^3 q}{ (2\pi)^{6} q} \Bigg[ [1/2 -\nf(\qk+q)]  \label{discDc} \\
\times &\Bigg\{ & \frac{s_3 s_4 [\hk_{+} \cdot \uqk_+] [\hk_{s_4} \cdot \uqk_{s_3}]
[1 - \nf(\qk) + n(q)]}{[ k + q + s_3\qk ][ q + \qk + s_4k] } \kd(E-k-\qk-q)
\nonumber \\
&-& \frac{s_3 s_4 [\hk_{-} \cdot \uqk_+] [\hk_{s_4} \cdot \uqk_{s_3}]
[1 - \nf(\qk) + n(q)]}{[ -k + q + s_3\qk ][ q + \qk + s_4k] } \kd(E+k-\qk-q) \Bigg\} 
\nonumber \\
&+&  [1/2 - \nf(q-\qk)]  \nonumber \\
\times &\Bigg\{&  - \frac{s_3 s_4 [\hk_{+} \cdot \uqk_-] [\hk_{s_4} \cdot \uqk_{s_3}]
[\nf(\qk) + n(q)]}{[ k + q + s_3\qk ][ -\qk + q + s_4k] } \kd(E-k+\qk-q)
\nonumber \\
&-&\frac{s_3 s_4 [\hk_{+} \cdot \uqk_+] [\hk_{s_4} \cdot \uqk_{s_3}]
[\nf(\qk) + n(q)]}{[ k - q + s_3\qk ][ \qk - q  + s_4k] } \kd(E-k-\qk+q)
\Bigg\} \Bigg]. \nonumber 
\end{eqnarray}


\section{The self-energy: Topology II}


\begin{figure}[htbp]
  \begin{center}
  \epsfxsize 80mm
  \epsfbox{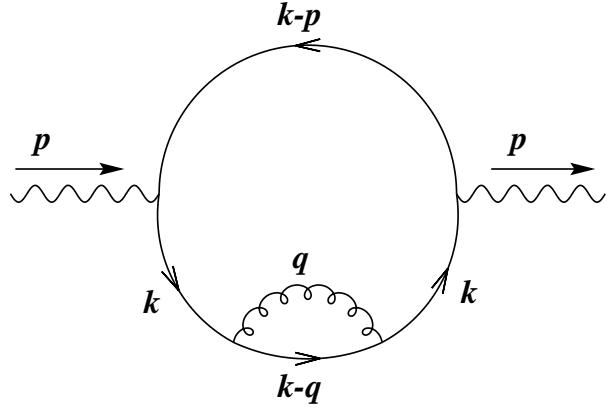}
\vspace{0.25cm}
    \caption{The second topology for the self-energy.}
    \label{othse}
  \end{center}
\end{figure}

We begin by evaluating the photon self energy with one quark 
line containing a gluon loop. The quark self energy may be written as:

\begin{equation}
-i\Sigma (k)_{i,k} = (ie)^2 \frac{i}{\beta} \sum_{q^0}  
\int \frac { d^{3}q }{ (2\pi)^{3} } t^{a}_{i,j} \gamma^{\mu} 
\frac{\f q}{q^2} t^{b}_{j,i} \gamma_{\mu} \frac{-i \kd^{a,b} }{(k-q)^2}.
\end{equation}

Using the identity $\gamma^{\mu}\f\!q \gamma_{\mu} = -2\f\!q$, 
$q$ being the momentum of the gluon, the Matsubara sum in the 
quark self energy is evaluated using the method of Pisarski 
\cite{pis88}. It is given in our notation(Appendix A) as 

\begin{equation}
\Sigma (k)_{i,k} = \frac{ g^{2} t^a_{i,j} t^b_{j,k} \kd^{a,b}}{ 2 } \int \frac { d^{3}q }{ (2\pi)^{3} } 
\sum_{ s_{1} s_{2} }
\frac { \wh{ (\f k - \f q) }_{s_{2}}  }{ q }
\frac { (s_{1}+s_{2})/2 - s_{1}\tilde{ n}_1 ( E_{k-q} )  + s_{2}n( q ) }
{ k^{0} - s_{2}E_{k-q} -s_{1}q } = \gamma^{\beta} {\Sigma_{i,k}}_{\beta}.
\end{equation}

Where, the self energy has been written in the 
final form to highlight its matrix structure. 
We may use this to write the full self energy of the 
photon in the static limit as 

\begin{eqnarray}
\Pi_{\mu}^{\mu} &=& \frac{1}{\beta} \sum_{k^{0}} \int  \frac{ d^3 k}{ (2\pi)^{3} }
(-1)Tr \sum_{ s_1 s_2 s_3 s_4 s_5 } \Bigg [ e \gamma_{\mu} 
\frac { \gamma^\alpha s_3 \hat{k}_{ \alpha , s_3 } \kd_{j,i}}{ 2(k^0 - s_3 k) }
 \gamma^{\beta} {\Sigma_{i,k}}_{\beta} (k) 
\frac { \gamma^\gamma s_4 \hat{k}_{ \gamma , s_4 } \kd_{k,l}}{ 2(k^0 - s_4 k) }
e \gamma^{\mu} \frac{ \gamma^\delta s_5 \hat{k}_{ \delta , s_5 } \kd_{l,j}}{ 2(k^0-p^0 - s_5 k) }
\Bigg].  
\end{eqnarray}
We choose the $z$ direction to be defined by the direction of $k$. Note that 
$\hk_{s_4} \cdot \hk_{s_5}= -2\delta_{s_4, -s_5}$. Note also that the $s_4$ 
and $s_3$ dependence of the
photon self-energy is identical. This allows us to write down the self-energy as,

\begin{equation}
\Pi_{\mu}^{\mu} = \frac{-1}{\beta} \sum_{k^{0}} \int  \frac{ d^3 k}{ (2\pi)^{3} }
2e^2 \Bigg[ 
\frac{ 2(-s_3) \Sigma_{jj} \cdot \hk_{s_3} }{( k^0 - s_3k )( k^0 - s_4k )( k^0-p^0 + s_4k )  }
- \frac{ 2 (-s_5)\Sigma_{jj} \cdot \hk_{s_5} }{( k^0-p^0 - s_5k )( (k^0)^2 - k^2 )} 
\Bigg].
\end{equation} 
Where summation is implied over all the sign variables present 
\ie, $s_1,s_2,s_3,s_4,s_5$. Note that the double pole 
is only present in the first term. 

We now need to evaluate the Matsubara sum over $k^0$. For this we follow the method of
reference \cite{kap89}. This method converts the Matsubara sum into a contour integration 
in the complex plain of $k^0$. The color factor from the quark self energy combined with 
that from the rest of the diagram becomes $\mbox{tr}[t^a,t^b] \kd^{ab} = 4$. Using this we 
obtain the self energy of the photon as:

\begin{eqnarray}
\Pi_{\mu}^{\mu} &=& \frac{8 e^2 g^2}{2 \pi i}  \int_{-i\infty + \epsilon}^{i\infty + \epsilon} dk^0
\int \frac{ d^3 k d^3 q}{ (2\pi)^{6} }
\frac {[ 1/2 - \nf(k^0) ][ (s_1+s_2)/2 - s_1\nf(\kq) + s_2 n(q) ]}{q [ sk^0 - s_2\kq -s_1q ]} 
\nonumber \\
&\times & \Bigg\{ \frac{ s_3 \wh{( k - q )}_{s_2} \cdot \hk_{s_3} }{[ sk^0 - s_3k ][ sk^0 - s_4k ]
[ sk^0 - p^0 + s_4k ] }  - 
\frac{ s_5 \wh{( k - q )}_{s_2} \cdot \hk_{s_5} }{[ sk^0 - p^0 - s_5k ][ (k^0)^2 -k^2 ]}
\Bigg\}.
\end{eqnarray}
The $k^0$ integration is from $-\infty \rightarrow \infty$ on the positive side of the real axis. 
We may
thus close the contour on the positive side. 
Note that the function is vanishing as $k^0 \rightarrow
\infty$. The result of this integration will simply be 
the sum of the residues at the corresponding
poles. Looking at the above expression we note the following poles:

\vspace{0.5cm}

i) 2nd order pole: at $k^0=k$, requires $s_3=s_4=s$ (only in the first term).

\vspace{0.5cm}

ii) 1st order pole: at $k^0=k$, no requirement (only in the second term); 
requires $s_3=-s_4=s$ or $-s_3=s_4=s$ (only in the first term).

\vspace{0.5cm}

iii) 1st order pole: at $k^0=k+sp^0$, requires $s_4=-s$ (only in first term); requires $s_5=s$
 (only in second term).

\vspace{0.5cm}

iv) 1st order pole: at $k^0 = s s_2 \kq + s s_1 q$, requires $s s_2 \kq + s s_1 q > 0$. 
 
\vspace{0.5cm}

In the following each of these poles will be evaluated in a separate subsection
 and then summed up.
 


\subsection{2nd order pole at $k^0=k$}


We begin by evaluating the 2nd order pole. The origin of this pole can be 
traced back to the two propagators whick may go on-shell simultaneously.
In the real time formalism this leads to the 
ill-defined square of the Dirac delta function. In imaginary time, however, this pole 
is easily dealt with: the residue of a function $f(k^0)$ at a second order 
pole at $k^0=k$ is simply given as $\frac{d}{dk^0} (k^0 - k)^2 f(k^0) |_{k^0=k}$. 
Using this we get the residue of $\Pi$ at this pole as

\begin{eqnarray}
\Pi_{\mu}^{\mu}(A) &=&  8 e^2 g^2 \int \frac{ d^3 k d^3 q}{ (2\pi)^{6} } 
\frac {[ (s_1+s_2)/2 - s_1\nf(\kq) + s_2 n(q) ]}{ q } 
\wh{( k - q )}_{s_2} \cdot \hk_{s}\nonumber \\
&\times & \Bigg\{ \frac{ s \onp }{ [sk - s_2\kq -s_1q ][ p^0 - 2sk ]} -
\frac{ 1/2 - \nf(k) }{ [sk - s_2\kq -s_1q ]^2[ p^0 - 2sk ]} \nonumber \\ 
& & \mbox{} + \frac{ 1/2 - \nf(k) }{ [sk - s_2\kq -s_1q ][ p^0 - 2sk ]^2} \Bigg\}.
\end{eqnarray} 

\noindent Note that as in the case of the first self-energy topology there 
is an extra negative sign in the residue as the countour is closed in the clockwise sense. 

\subsection{1st order pole at $k^0=k$}

This obvious residue may be easily evaluated using the methods outlined in the 
previous section, 

\begin{eqnarray}
\Pi_{\mu}^{\mu}(B) &=& -8 e^2 g^2 \int \frac{ d^3 k d^3 q}{ (2\pi)^{6} } 
\frac {[ (s_1+s_2)/2 - s_1\nf(\kq) + s_2 n(q) ][ 1/2 - \nf(k) ]}{ q [sk - s_2 \kq - s_1 q]}
 \nonumber \\
&\times & \Bigg\{  \frac { -s \wh{( k - q )}_{s_2} \cdot \hk_{s} }{2kp^0} +
\frac { s \wh{( k - q )}_{s_2} \cdot \hk_{-s} }{2k[p^0-2sk]} 
+ \frac { s_5 \wh{( k - q )}_{s_2} \cdot \hk_{s_5} }{2k[p^0-(s-s_5)k]} 
\Bigg\}.
\end{eqnarray}

\noindent In the above, we sum over the two possibilities of $s_5 = \pm s$ to get 
the factor in the bracket as

\begin{eqnarray*}
\Bigg\{  \frac { -s \wh{( k - q )}_{s_2} \cdot \hk_{s} }{2kp^0} +
\frac { s \wh{( k - q )}_{s_2} \cdot \hk_{-s} }{2k[p^0-2sk]} 
+ \frac { s \wh{( k - q )}_{s_2} \cdot \hk_{s} }{2kp^0} \\
+ \frac { -s \wh{( k - q )}_{s_2} \cdot \hk_{-s} }{2k[p^0-2sk]}
\Bigg\} = 0
\end{eqnarray*} 

\noindent Hence, 
$\Pi_{\mu}^{\mu}(B) = 0$.


\subsection{1st order pole at $k^0=k+sp^0$}


This gives the residue
 
\begin{eqnarray}
\Pi_{\mu}^{\mu}(C) &=& -8 e^2 g^2 \int \frac{ d^3 k d^3 q}{ (2\pi)^{6} } 
\frac {[ (s_1+s_2)/2 - s_1\nf(\kq) + s_2 n(q) ][ 1/2 - \nf(k) ]}{ q [p^0 -( s_2 \kq + s_1 q - sk )]
[ p^0 + 2sk ]} 
\nonumber \\
&\times & \Bigg\{ \frac{ s_3 \wh{( k - q )}_{s_2} \cdot \hk_{s_3} }{s(p^0 -( s_3 - s )k) } -
\frac {\wh{( k - q )}_{s_2} \cdot \hk_{s} }{p^0}
\Bigg\}.
\end{eqnarray} 

\noindent Switching $s \ra -s$, and summing over $s_3 = \pm s $ we get,

\begin{eqnarray}
\Pi_{\mu}^{\mu}(C) &=& 8 e^2 g^2 \int \frac{ d^3 k d^3 q}{ (2\pi)^{6} } 
\frac {[ (s_1+s_2)/2 - s_1\nf(\kq) + s_2 n(q) ][ 1/2 - \nf(k) ]}{ q [p^0 -( s_2 \kq + s_1 q + sk )]
[ p^0 - 2sk ]} 
\nonumber \\
&\times & \Bigg\{ \frac{ \wh{( k - q )}_{s_2} \cdot \hk_{s} }{( p^0 - 2sk) } \Bigg\}.
\end{eqnarray}


\subsection{1st order pole at $k^0= s s_2 \kq + s s_1 q$}


This pole is realized only if $s s_2 \kq + s s_1 q > 0$. This condition may be enforced with the
following set of delta and theta functions:

\begin{equation}
\kd_{s,s_2} \kd_{s,s_1} + \kd_{s,s_2} \kd_{s,-s_1} \Theta (\kq - q) 
+ \kd_{s,-s_2} \kd_{s,s_1} \Theta (q-\kq). \label{theta=}
\end{equation}

We start with the second and third terms:

\begin{eqnarray}
\Pi_{\mu}^{\mu}(D,2) &=& 8 e^2 g^2 \int \frac{ d^3 k d^3 q}{ (2\pi)^{6} } 
\frac{[ 1/2 - \nf( \kq - q )][ \nf(\kq) + n(q) ]}{q} \nonumber \\
&\times& \Bigg\{ 
\frac{ s_3 \wh{( k - q )}_{s} \cdot \hk_{s_3} }
{[s\kq - sq - s_3k][s\kq - sq - s_4k][p^0 - ( s_4k + s\kq - sq )]} \nonumber \\
& & \mbox{} - \frac { s_5 \wh{( k - q )}_{s} \cdot \hk_{s_5}  }
{[p^0 - ( s\kq - sq - s_5k )][(\kq-q)^2-k^2]}
\Bigg\} \Theta(\kq - q).
\end{eqnarray}


\noindent Similarly we find for the third term 

\begin{eqnarray}
\Pi_{\mu}^{\mu}(D,3) &=& 8 e^2 g^2 \int \frac{ d^3 k d^3 q}{ (2\pi)^{6} } 
\frac{[ 1/2 - \nf( q - \kq )][ - \nf(\kq) - n(q) ]}{q} \nonumber \\
&\times& \Bigg\{ 
\frac{ s_3 \wh{( k - q )}_{-s} \cdot \hk_{s_3} }
{[-s\kq + sq - s_3k][-s\kq + sq - s_4k][p^0 - ( s_4k - s\kq + sq )]} \nonumber \\
& & \mbox{} - \frac { s_5 \wh{( k - q )}_{-s} \cdot \hk_{s_5}  }
{[p^0 - ( -s\kq + sq - s_5k )][(\kq-q)^2-k^2]}
\Bigg\} \Theta( q - \kq ). 
\end{eqnarray}

\noindent Now, switch $s \ra -s$ in the 3rd term, and noting that 
$1/2 - \nf(q-\kq) = -1/2 + \nf(\kq -q)$, we
note that the second and third terms can be combined to give 

\begin{eqnarray}
\Pi_{\mu}^{\mu}(D,2+3) &=& 8 e^2 g^2 \int \frac{ d^3 k d^3 q}{ (2\pi)^{6} } 
\frac{[ 1/2 - \nf( \kq - q )][ \nf(\kq) + n(q) ]}{q} \nonumber \\
&\times& \Bigg\{ 
\frac{ s_3 \wh{( k - q )}_{s} \cdot \hk_{s_3} }
{[s\kq - sq - s_3k][s\kq - sq - s_4k][p^0 - ( s_4k + s\kq - sq )]} \nonumber \\
& & \mbox{} - \frac {s_5 \wh{( k - q )}_{s} \cdot \hk_{s_5}  }
{[p^0 - ( s\kq - sq - s_5k )][(\kq-q)^2-k^2]}
\Bigg\}. 
\end{eqnarray}


\noindent Note the absence of the theta functions in the above equation. 
Now we may also write down the residue
from the first set of delta functions in Eq.~(\ref{theta=}) as,

\begin{eqnarray}
\Pi_{\mu}^{\mu}(D,1) &=& 8 e^2 g^2 \int \frac{ d^3 k d^3 q}{ (2\pi)^{6} } 
\frac{[ 1/2 - \nf( \kq + q )][1-\nf(\kq) + n(q) ]}{q} \nonumber \\
&\times& \Bigg\{ 
 \frac{ s_3 \wh{( k - q )}_{s} \cdot \hk_{s_3} }
{[s\kq + sq - s_3k][s\kq + sq - s_4k][p^0 - ( s_4k + s\kq + sq )]} \nonumber \\
& & \mbox{} - \frac {s_5 \wh{( k - q )}_{s} \cdot \hk_{s_5}  }
{[p^0 - ( s\kq + sq - s_5k )][(\kq+q)^2-k^2]}
\Bigg\}. 
\end{eqnarray}


 The total expression obtained by summing up the results from the preceding 4 subsections will give
us the full self-energy of the photon to second order in the coupling constant for the diagram of 
Fig.~\ref{othse}, \ie,

\[
\Pi_{\mu}^{\mu} = \Pi_{\mu}^{\mu}(A) + \Pi_{\mu}^{\mu}(B) + \Pi_{\mu}^{\mu}(C) 
+ \Pi_{\mu}^{\mu}(D,1) + \Pi_{\mu}^{\mu}(D,2+3).
\]


\section{IMAGINARY PART OF THE SECOND SELF-ENERGY topology}


We now proceed with evaluating the discontinuity in 
the second self-energy as $p^0$ is analytically
continued to a positive real value \ie, 
$p^0 \ra E + i\epsilon$. Analyzing the expressions derived
in the above sections we note the following discontinuities:

\vspace{0.5cm}

a) Poles of type $p^0=2k$: 

\vspace{0.5cm}

i) 1st order pole in $\Pi_{\mu}^{\mu}(A)$: 
at $p^0=2k$, requires $s=1$ (only in first and second terms).

\vspace{0.5cm}

ii) 2st order pole in $\Pi_{\mu}^{\mu}(A)$: 
at $p^0=2k$. This occurs in the third term in the bracket and requires $s=1$.

\vspace{0.5cm}

iii) 2nd order pole in $\Pi_{\mu}^{\mu}(C)$:
 at $p^0=2k$, requires $s=1$ and $s_3=1$ (only in first 
 term).
 
\vspace{0.5cm}

b) Poles of type $p^0=sk+s_1q+s_2\kq$:

\vspace{0.5cm}

iv) 1st order pole in $\Pi_{\mu}^{\mu}(C)$: at 
$p^0=sk+s_1q+s_2\kq$, requires $s=s_1=s_2=1$, or 
$-s=s_1=s_2=1$, or $s=-s_1=s_2=1$, or $s=s_1=-s_2=1$ (in both terms ).

\vspace{0.5cm}

v) 1st order pole in $\Pi_{\mu}^{\mu}(D,2+3)$: at 
$p^0=s_4k+s\kq-sq$, requires $s_4=s=1$ or
$s_4=-s=1$ (only in first term); at $p^0=-s_5k+s\kq-sq$, 
requires $-s_5=s=1$ or $s_5=s=-1$
(only in second term).

\vspace{0.5cm}

vi) 1st order pole in $\Pi_{\mu}^{\mu}(D,1)$: at 
$p^0=s_4k+s\kq+sq$, requires $s_4=s=1$ or
$-s_4=s=1$ (only in first term), at  $p^0=-s_5k+s\kq+sq$, 
requires $-s_5=s=1$ or $s_5=s=1$
(only in second term).

The discontinuity across a second order pole is derived in Appendix B.
We now write down the various discontinuities as enumerated above:

\begin{eqnarray}
Disc[\Pi_{\mu}^{\mu}(A)]_a &=& (-2\pi i) 8 e^2 g^2 \int \frac{ dk d\h d\phi \sin{\h} d^3 q}
{ (2\pi)^{6} q}  \kd(E-2k) 
\nonumber \\
&\times& \Bigg\{ \frac { k^2 \N \Sc  [ 1/2 - \nf(k) ]\p }{ [ k - s_2\kq - s_1q ]} 
-\frac{ k^2 \N \Sc  [ 1/2 - \nf(k) ] }{ [ k - s_2\kq -s_1q ]^2 } \nonumber \\
& & - \frac{1}{2} \frac { 2k \N \Sc  [ 1/2 - \nf(k) ] }{ [ k - s_2\kq -s_1q ]}
- \frac{1}{2} \frac { k^2 (\N \Sc )\p  [ 1/2 - \nf(k) ] }{ [ k - s_2\kq -s_1q ]} \nonumber \\
& & - \frac{1}{2} \frac { k^2 \N \Sc  [ 1/2 - \nf(k) ]\p }{ [ k - s_2\kq - s_1q ]}
- \frac{1}{2} \frac{ k^2 \N \Sc  [ 1/2 - \nf(k) ] [ -1 + s_2 \kq\p ]}{ [ k - s_2\kq -s_1q ]^2 } 
\Bigg\}. \label{discA}
\end{eqnarray}

\noindent Where, the prime ``$A\p$''  denotes derivation only with respect to $k$. 
The symbol $\N$ stands for the
factor $[ (s_1+s_2)/2 - s_1\nf(\kq) + s_2 n(q) ]$, while the factor 
$\wh{( k - q )}_{s_2} \cdot \hk_{+}$ is represented by the symbol $\Sc$.

\begin{eqnarray}
Disc[\Pi_{\mu}^{\mu}(C)]_a &=& (2 \pi i) 8 e^2 g^2 \int \frac{ dk d\h d\phi \sin{\h} d^3 q}
{ (2\pi)^{6} }  \kd(E-2k) 
\nonumber \\
&\times&  \Bigg\{ \frac{1}{2} \frac { 2k \N \Sc  [ 1/2 - \nf(k) ] }{ [ E - s_2\kq - s_1q - k ]}
+ \frac{1}{2} \frac { k^2 (\N \Sc )\p  [ 1/2 - \nf(k) ] }{ [ E - s_2\kq - s_1q - k ]} \nonumber \\
& & + \frac{1}{2} \frac { k^2 \N \Sc  [ 1/2 - \nf(k) ]\p }{ [ E - s_2\kq - s_1q - k ]}
+ \frac{1}{2} \frac{ k^2 \N \Sc  [ 1/2 - \nf(k) ] [ 1 + s_2 \kq\p ]}{ [ E - s_2\kq -s_1q - k]^2 } 
\Bigg\}.\label{discCa}
\end{eqnarray}

The two terms above are the result of the discontinuities at $p^0 \ra E = 2k$. 
In the following we
shall enumerate those terms that result as we take the discontinuities 
at $p^0 \ra E = sk+s_1q+s_2\kq$.

\begin{eqnarray}
Disc[\Pi_{\mu}^{\mu}(C)]_c &=& (-2\pi i) 8 e^2 g^2 \int \frac{ d^3 k d^3 q}{ (2\pi)^{6} } 
\frac {[ (s_1+s_2)/2 - s_1\nf(\kq) + s_2 n(q) ][ 1/2 - \nf(k) ]}{ q 
[  s_2 \kq + s_1 q - sk ]} 
\nonumber \\
&\times& \Bigg\{ \frac{s_3 \wh{( k - q )}_{s_2} \cdot \hk_{s_3} }{ s[  s_2 \kq + s_1 q - s_3k ]} -
\frac {\wh{( k - q )}_{s_2} \cdot \hk_{-s} }{ s_2 \kq + s_1 q + sk }
\Bigg\} \kd (E - sk - s_1 q - s_2 \kq )  \nonumber \\
&\times& \Big[ \kd_{s,+} \kd_{s_1,+} \kd_{s_2,+} +  \kd_{s,-} \kd_{s_1,+} \kd_{s_2,+} + 
 \kd_{s,+} \kd_{s_1,-} \kd_{s_2,+} 
+ \kd_{s,+} \kd_{s_1,+} \kd_{s_2,-}  \Big].
\end{eqnarray}
 
\noindent Recall that even though not explicitly mentioned there is 
an implied summation over all sign factors.
We may now perform the sum over $s_3 = \pm s$ to get 

\begin{eqnarray}
Disc[\Pi_{\mu}^{\mu}(C)]_c &=& (-2\pi i) 8 e^2 g^2 \int \frac{ d^3 k d^3 q}{ (2\pi)^{6} } 
\frac {[ (s_1+s_2)/2 - s_1\nf(\kq) + s_2 n(q) ][ 1/2 - \nf(k) ]}{ q 
[  s_2 \kq + s_1 q - sk ]^2} 
\nonumber \\
&\times&  \wh{( k - q )}_{s_2} \cdot \hk_{s} 
 \kd (E - sk - s_1 q - s_2 \kq )  \nonumber \\
&\times& \Big[ \kd_{s,+} \kd_{s_1,+} \kd_{s_2,+} +  \kd_{s,-} \kd_{s_1,+} \kd_{s_2,+} + 
 \kd_{s,+} \kd_{s_1,-} \kd_{s_2,+} 
+ \kd_{s,+} \kd_{s_1,+} \kd_{s_2,-}  \Big]. \label{discCc}
\end{eqnarray}

\noindent Finally, the discontinuity in parts D of the second self energy is given as

\begin{eqnarray}
Disc[\Pi_{\mu}^{\mu}(D,2+3)]_c &=& (-2 \pi i) 8 e^2 g^2 \int \frac{ d^3 k d^3 q}{ (2\pi)^{6} } 
\frac{[ 1/2 - \nf( \kq - q )][ \nf(\kq) + n(q) ]}{q} \nonumber \\
& & \!\!\!\!\!\!\!\!\!\!\!\!\!\!\!\!\!\!\!\!\!\!\!\!\!\!\!\!\!\!\!\!\!\!\!\!\!\!\!\!\!\!\!\!\!\!\!\!
\!\!\!\!\!\!\!\!\!\!\!\!\!\!\!\!\!\!\!\!
\Bigg\{ 
\frac{ s_3 \wh{( k - q )}_{s} \cdot \hk_{s_3} }
{[s\kq - sq - s_3k][s\kq - sq - s_4k]} \Big[ \kd_{s_4,+}\kd_{s,+}\kd(E-k-\kq+q) 
+ \kd_{s_4,+}\kd_{s,-}\kd(E-k+\kq-q) \Big] \nonumber \\
& & \!\!\!\!\!\!\!\!\!\!\!\!\!\!\!\!\!\!\!\!\!\!\!\!\!\!\!\!\!\!\!\!\!\!\!\!\!\!\!\!\!\!\!\!\!\!\!\!
\!\!\!\!\!\!\!\!\!\!\!\!\!\!\!\!\!\!\!\!
\mbox{} - \frac { s_5 \wh{( k - q )}_{s} \cdot \hk_{s_5}  }
{[(\kq-q)^2-k^2]} \Big[ 
\kd_{s_5,-}\kd_{s,+}\kd(E-k-\kq+q) 
+ \kd_{s_5,-}\kd_{s,-}\kd(E-k+\kq-q) 
\Big]
\Bigg\}.
\end{eqnarray}

\noindent We may now sum over $s_3 = \pm 1$ to get

\begin{eqnarray}
Disc[\Pi_{\mu}^{\mu}(D,2+3)]_c &=& (-2 \pi i) 8 e^2 g^2 \int \frac{ d^3 k d^3 q}{ (2\pi)^{6} } 
\frac{[ 1/2 - \nf( \kq - q )][ \nf(\kq) + n(q) ]}{q} \nonumber \\
& & \!\!\!\!\!\!\!\!\!\!\!\!\!\!\!\!\!\!\!\!\!\!\!\!\!\!\!\!\!\!\!\!\!\!\!\!\!\!\!\!\!\!\!\!\!\!\!\!
\!\!\!\!\!\!\!\!\!\!\!\!\!\!\!\!\!\!\!\!
\times \Bigg\{ \frac{ \wh{(k-q)}_- \cdot \hk_+ }{ [ k + \kq - q ]^2 } \kd(E-k+\kq-q) 
+ \frac{ \wh{(k-q)}_+ \cdot \hk_+ }{ [ k + q - \kq ]^2 } \kd(E-k-\kq+q)
\Bigg\}. \label{discD2+3c}
\end{eqnarray}

\begin{eqnarray}
Disc[\Pi_{\mu}^{\mu}(D,1)]_c &=& (-2 \pi i) 8 e^2 g^2 \int \frac{ d^3 k d^3 q}{ (2\pi)^{6} } 
\frac{[ 1/2 - \nf( \kq + q )][1-\nf(\kq) + n(q) ]}{q} \nonumber \\
& &\!\!\!\!\!\!\!\!\!\!\!\!\!\!\!\!\!\!\!\!\!\!\!\!\!\!\!\!\!\!\!\!\!\!\!\!\!\!\!\!\!\!\!\!\!\!\!\!
\!\!\!\!\!\!\!\!\!\! 
\times \Bigg\{ 
\frac{ s_3 \wh{( k - q )}_{s} \cdot \hk_{s_3} }
{[s\kq + sq - s_3k][s\kq + sq - s_4k]} \Big[ \kd_{s_4,+}\kd_{s,+}\kd(E-k-\kq-q) 
+ \kd_{s_4,-}\kd_{s,+}\kd(E+k-\kq-q) \Big] \nonumber \\
& & \!\!\!\!\!\!\!\!\!\!\!\!\!\!\!\!\!\!\!\!\!\!\!\!\!\!\!\!\!\!\!\!\!\!\!\!\!\!\!\!\!\!\!\!\!\!\!\!
\!\!\!\!\!\!\!\!\!\!
\mbox{} - \frac { s_5 \wh{( k - q )}_{s} \cdot \hk_{s_5}  }
{[(\kq+q)^2-k^2]}\Big[ 
\kd_{s_5,-}\kd_{s,+}\kd(E-k-\kq-q) 
+ \kd_{s_5,+}\kd_{s,+}\kd(E+k-\kq-q) 
\Big]
\Bigg\} .
\end{eqnarray}

\noindent We may sum over $s_3 = \pm 1$ to obtain 

\begin{eqnarray}
Disc[\Pi_{\mu}^{\mu}(D,1)]_c &=& (-2 \pi i) 8 e^2 g^2 \int \frac{ d^3 k d^3 q}{ (2\pi)^{6} } 
\frac{[ 1/2 - \nf( \kq + q )][1-\nf(\kq) + n(q) ]}{q} \nonumber \\
& & \!\!\!\!\!\!\!\!\!\!\!\!\!\!\!\!\!\!\!\!\!\!\!\!\!\!\!\!\!\!\!\!\!\!\!\!\!\!\!\!\!\!\!\!\!\!\!
\!\!\!\!
\times \Bigg\{ \frac{\wh{(k-q)}_+ \cdot \hk_+ }{ [ \kq + q - k ]^2 } \kd(E-k-\kq-q)
- \frac{\wh{(k-q)}_+ \cdot \hk_- }{ [ \kq + q + k ]^2 } \kd(E+k-\kq-q)
\Bigg\} .\label{discD1c}
\end{eqnarray}


\section{PHYSICAL INTERPRETATION: TREE-LIKE CUTS}


We now begin the process of combining terms from the discontinuities of the two self energies 
to obtain the square of amplitudes of physical processes.  
Essentially we shall follow 
the method outlined by H. A. Weldon \cite{wel83}. Our method is essentially a 
three step process:

\noindent i) collect together terms that have the same energy conserving delta functions.

\noindent ii) reorganize the thermal distribution functions to express them as a difference 
of the thermal weights for particle emission and absorption.

\noindent iii) reorganize the remaining momentum dependent part as the square of the 
amplitude of the process hinted at by the previous two steps.

For easy identification we indicate the contribution from the 
first self-energy topology by $\Pi^1$ and from the second topology 
by $\Pi^2$.
We begin with the discontinuities where no loops are left in the final result. 
These are the discontinuities given by Eqs.~(\ref{discAc},\ref{discDc}) 
for the 
first self-energy topology, 
and Eqs.~(\ref{discCc},\ref{discD2+3c},\ref{discD1c}) for 
the second self-energy topology. 
These discontinuities will result in physical amplitudes 
for three kinds of processes: 
photon decay, Compton scattering and pair creation.  


\subsection{Photon Decay and Formation.}


We begin by analyzing the terms which containing the delta function $\kd(E-k-q-\kq)$. 
The contributions to this from \po are 

\begin{eqnarray} 
Disc[\Pi^{1}(A)](E-k-\qk-q) &=& (+2\pi i) 8 e^2 g^2  
\int \frac{ d^3 k d^3 q}{ (2\pi)^{6} q}  [1/2 - \nf(k)]  \kd(E-k-\qk-q) \nonumber \\
&\times  & \frac{s_3 s_4 [\hk_{+} \cdot \uqk_+] [\hk_{s_4} \cdot \uqk_{s_3}]
[-1 + \nf(\qk) - n(q)]}{[ k + q + s_3\qk ][ q + \qk + s_4k] }.
\end{eqnarray}

\noindent and 

\begin{eqnarray} 
Disc[\Pi^{1}(D)](E-k-\qk-q) &=& (-2\pi i) 8 e^2 g^2  
\int \frac{ d^3 k d^3 q}{ (2\pi)^{6} q}  [1/2 -\nf(\qk+q)] \kd(E-k-\qk-q) \nonumber \\
&\times  & \frac{s_3 s_4 [\hk_{+} \cdot \uqk_+] [\hk_{s_4} \cdot \uqk_{s_3}]
[1 - \nf(\qk) + n(q)]}{[ k + q + s_3\qk ][ q + \qk + s_4k] }.
\end{eqnarray} 

Note that

\[
\nf(\qk+q) \Big[ 
1 - \nf(\qk) + n(q) \Big] =  \nf(\qk)n(q)
\]

using the above identity we may combine the two terms and rewrite the 
distribution functions to give,

\begin{eqnarray} 
Disc[\Pi^{1}](E-k-\qk-q) &=& (-2\pi i) 8 e^2 g^2  
\int \frac{ d^3 k d^3 q}{ (2\pi)^{6} q}  \nonumber \\
&\times  & \Big\{ [ 1 - \nf(k) ][ 1 + n(q) ][ 1 - \nf(\qk) ] -  \nf(k)n(q)\nf(\qk) \Big\}
\nonumber \\
&\times  & \frac{s_3 s_4 [\hk_{+} \cdot \uqk_+] [\hk_{s_4} \cdot \uqk_{s_3}]}
{[ k + q + s_3\qk ][ q + \qk + s_4k] } \kd(E-k-\qk-q).
\end{eqnarray}

We may combine the coefficients of the same delta function from the second self-energy to get 

\begin{eqnarray}
Disc[\Pi^{2}](E-k-q-\kq) &=& 2 \times (-2 \pi i) 8 e^2 g^2 \int \frac{ d^3 k d^3 q}{ (2\pi)^{6} q} 
\nonumber \\
&\times  & \Big\{ [ 1 - \nf(k) ][ 1 + n(q) ][ 1 - \nf(\qk) ] -  \nf(k)n(q)\nf(\qk) \Big\}
\nonumber \\
& \times & \frac{\wh{(k-q)}_+ \cdot \hk_+ }{ [ \kq + q - k ]^2 } \kd(E-k-\kq-q). \label{phtndcy2}
\end{eqnarray}

\begin{figure}[htbp]
  \begin{center}
  \epsfxsize 120mm
  \epsfbox{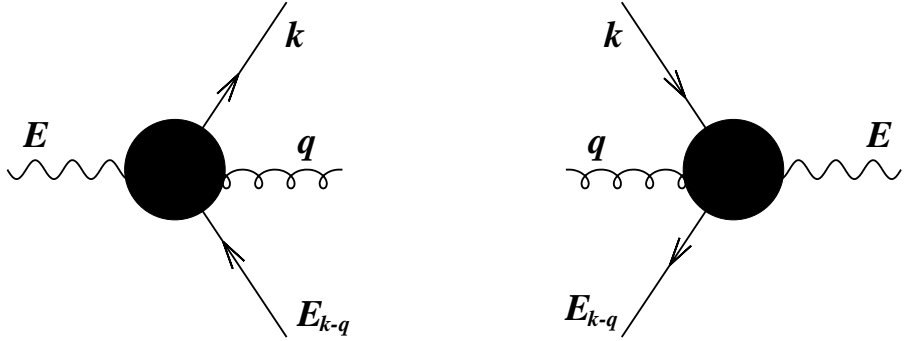}
\vspace{0.25cm}
    \caption{Heavy photon decay and formation.}
    \label{phtndecay}
  \end{center}
\end{figure}

The overall factor of 2 is the ratio of the symmetry factor of this diagram 
to the denominator 
obtained from perturbation theory. Note that we obtain the same form of the 
distribution functions, 
this indicates the generic structure of heavy photon decay and reformation. 
In the distribution 
function factor, terms like $1+n(q)$ indicate Bose-Einstein enhancement in 
the emission of a gluon. 
The $1$ is from spontaneous emission, and the $n(q)$ represents stimulated 
emission of a boson into 
a thermal bath. Terms like $1-\nf(k)$ represents the ``Pauli blocked'' 
emission of a quark of 
momentum $k$ into the thermal bath. The product of the three factors 
$[ 1 - \nf(k) ][ 1 + n(q) ][ 1 - \nf(\qk) ]$ (along with the phase 
space integral and delta function) can thus be interpreted as the 
statistical factor associated with a heavy photon outside a thermal 
bath decaying by emitting a quark, antiquark, and a gluon into a 
thermal bath. Subtracted from this is the factor $\nf(k)n(q)\nf(\qk)$; 
this represents the formation of a heavy photon from a quark, antiquark 
and a gluon, all three emitted from the thermal bath 
(Fig.~\ref{phtndecay}). The photon subsequently escapes from the bath 
without further interaction.   

To convert the above expressions into cross sections for heavy photon 
decay and reformation, we start by first defining a new four-vector 
$  \fw = (w,\vec{w})$, such that $w = E - k - q$(in order to avoid 
confusion we introduce the notation of four-vectors as bold face 
characters). This relation is indicated by the one dimensional delta 
function. To obtain the probability of photon decay, we need to 
generalize the delta function to a four-delta function. We thus need 
to generalize the definition of $\fw$:  

\[
\fw = \fp - \fk - \fq.
\] 

\noindent Where, $\fp = (E,0,0,0)$ is the mass of the off-shell photon. As 
denoted by Fig.~\ref{phtndecay}. $\fk$, $\fq$ and $\fw$ are all 
on shell. The above relation also implies $\vec{w} = 
- \vec{k} - \vec{q} $. We may set 
$
\fk^0 = k = |\vec{k}| $ and
$
\fq^0 = q = |\vec{q}| 
$.
Now, requiring that $\fw$ be on shell imposes the condition that 

\[
(E-k-q)^2 = k^2 + q^2 +2kq\cos{\h} 
\]

\[
\im E(E-2k-2q) = - \fk \cdot \fq. 
\]

\noindent Where $\h$ is the angle between the three vectors $\vq$ and $\vk$. 
Using the above relations we many now rewrite the discontinuity 
obtained from \pt. In the numerator of the integrand we notice 
the factor $\wh{(k-q)}_+ \cdot \hk_+$, this may be changed 
appropriately by setting $\vk \lra -\vk$ in the integrand. 
Noting that $\wh{(-k)}_s = -\hk_{-s}$ we get the above factor 
as  $-\hat{w}_+ \cdot \hk_-$. Introducing the standard 
denominators $2k,2w$ and factors of $2\pi$ we obtain 
Eq.(\ref{phtndcy2}) as

\begin{eqnarray}
Disc[\Pi^{2}](E-k-q-\kq) &=& -2i \times 8 e^2 g^2 \int 
\frac{ d^3 k d^3 q d^3 w}{ (2\pi)^{9} 2q 2k 2w} 8 (2 \pi )^4 \kd^4(\fp-\fk-\fq-\fw)
\nonumber \\
& \times & \Big\{ [ 1 - \nf(k) ][ 1 + n(q) ][ 1 - \nf(w) ] -  \nf(k)n(q)\nf(w) \Big\}
\nonumber \\
& \times & \Bigg[ -\frac{ \fw_+ \cdot \fk_- }{ [ w + q - k ]^2 } \Bigg] 
\end{eqnarray}
 
We now split the above integrand into two parts and in one of them switch 
$w \leftrightarrow k$. Note that 
$- \fw_+ \cdot \fk_- = wk + \vw \cdot \vk = (E-2k)(E-2w)/2$, 
finally gives the above discontinuity as

\begin{eqnarray}
Disc[\Pi^{2}](E-k-q-\kq) &=&  -i \int \frac{ d^3 k d^3 q d^3 w }{ (2\pi)^{9} 2q 2k 2w }
(2 \pi )^4 \kd^4( \fp - \fk - \fq - \fw )
\nonumber \\
&\times& \Big\{ [ 1 - \nf(k) ][ 1 + n(q) ][ 1 - \nf(w) ] -  \nf(k) n(q) \nf(w) \Big\}
\nonumber \\
&\times& 32 e^2 g^2 \Bigg[ \frac{ E - 2k }{ E - 2w } + \frac{ E - 2w }{ E - 2k } \Bigg]. \label{lp2phtndcy}
\end{eqnarray}

\noindent We now perform the same procedure on the corresponding discontinuity from \po, to get

\begin{eqnarray}
Disc[\Pi^{1}](E-k-q-\kq) &=& -i \times 8 e^2 g^2 \int 
\frac{ d^3 k d^3 q d^3 w}{ (2\pi)^{9} 2q 2k 2w} 8 (2 \pi )^4 \kd^4(\fp-\fk-\fq-\fw)
\nonumber \\
&\times& \Big\{ [ 1 - \nf(k) ][ 1 + n(q) ][ 1 - \nf(w) ] -  \nf(k)n(q)\nf(w) \Big\}
\nonumber \\
&\times& \frac{ s_3 s_4 [ \fk_{+} \cdot \fw_+ ] [ \hk_{s_4} \cdot \hw_{s_3} ]}
{[ k + q + s_3 w ][ q + w + s_4 k ] } .
\end{eqnarray} 

\noindent The part of the integrand besides the distribution function 
part(depends on the angle between $\vk$ and $\vq$, 
will be denoted as the matrix part) may be expanded by summing over $s_3,s_4$ as

\[
\frac{ \fk \cdot \fw }{kw} \left[ \frac{ \fk_+ \cdot \fw_+ }{ (k+q+w)^2 } + 
\frac{ \fk_+ \cdot \fw_- }{ (k+q)^2 - w^2 } + 
\frac{ \fk_- \cdot \fw_+ }{ (q+w)^2 - k^2 } + 
\frac{ \fk_- \cdot \fw_- }{ ( q^2 - (k-w)^2 } \right].
\]

Using the relations

\[
\fk_+ \cdot \fw_- = \fk_- \cdot \fw_+ = -(1/2) [E-2k][E-2w],
\]

\[
\fk_+ \cdot \fw_+ = \fk_- \cdot \fw_- = (E/2)[E-2q],
\]

\noindent and the relation imposed by the delta function (\ie,$E=k+q+w$) we can simplify 
the matrix part to give 

\[
\frac{E(E-2q)}{[E-2w][E-2k]},
\]

\noindent substituting the above into the expression for \po and then combining 
the results from \po and \pt we get

\begin{eqnarray}
Disc[\Pi](E-k-q-\kq) &=&  -i \int \frac{ d^3 k d^3 q d^3 w }{ (2\pi)^{9} 2q 2k 2w }
(2 \pi )^4 \kd^4( \fp - \fk - \fq - \fw )
\nonumber \\
&\times & \Big\{ [ 1 - \nf(k) ][ 1 + n(q) ][ 1 - \nf(w) ] -  \nf(k) n(q) \nf(w) \Big\}
\nonumber \\
&\times & 32 e^2 g^2 \Bigg[ \frac{ E - 2k }{ E - 2w } + \frac{ E - 2w }{ E - 2k }  + 
2 \frac{E(E-2q)}{[E-2w][E-2k]} \Bigg]. \label{lpphtndcy}
\end{eqnarray}

\begin{figure}[htbp]
  \begin{center}
  \epsfxsize 120mm
  \epsfbox{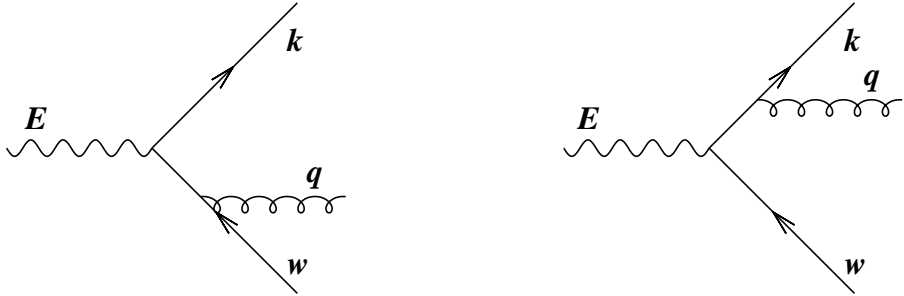}
\vspace{0.25cm}
    \caption{Heavy photon decay at first order in $\alpha$ and $\alpha_s$.}
    \label{phtndecay1}
  \end{center}
\end{figure}

Photon decay into a quark, antiquark and a gluon at first order in the E.M. 
and strong coupling constant can occur by two types of Feynman diagrams 
\cite{fie89} as shown in Fig.~\ref{phtndecay1}.
The matrix element for the first diagram may be written as 
$\mat_{1}=\mat_{1}^{\mu} \epsilon _{\mu}(\fp)$, where 

\[
\mat_{1}^{\mu} = \bar{u}(k) i e \gamma ^{\mu} \frac{i( \f \fk - \f \fp )}{(k-p)^2} ig 
\gamma ^{\rho} \epsilon^{*}_{\rho} (q) v(w)
\]

\noindent for the second diagram 

\[
\mat_{2}^{\mu} = \bar{u}(k) ig \gamma ^{\rho} \epsilon^{*}_{\rho}(q) 
\frac{i( \f \fp - \f \fw )}{(p-w)^2} i e \gamma ^{\mu} v(w)
\]

\noindent Taking the product ${\mat_{1}^*}^{\mu}{\mat_{1}}_{\mu}$ and summing over the 
spins and colors of the quark, antiquark and the gluon gives

\begin{equation}
{\mat^{*}_{1}}^{\mu}{\mat_1}_{\mu} = -32 e^2 g^2 \frac{E-2w}{E-2k} . 
\end{equation}

\noindent Similarly 

\begin{equation}
{\mat^{*}_2}^{\mu}{\mat_2}_{\mu} = -32 e^2 g^2 \frac{E-2k}{E-2w} . 
\end{equation}

Notice that as the three 3-vectors $\vk,\vq,\vw$ form a 
triangle, $E-2k= w+q-k$ is always positive. By the same 
argument $E-2w$ and $E-2q$ are also positive. We thus note 
that ${\mat^{*}_{1}}^{\mu}{\mat_1}_{\mu},{\mat^{*}_2}^{\mu}{\mat_2}_{\mu}$ 
are negative. This is to be expected as the square of the full matrix 
element $|\mat|^2$ is positive, where from the sum over the photon's spin we get 

\[
|\mat|^2 = -g_{\mu,\nu} {\mat^{*}}^{\mu} {\mat}^{\nu}
= -{\mat^{*}}^{\mu} {\mat}_{\mu} .
\]

\noindent The cross term is 

\begin{equation}
{\mat^{*}_2}^{\mu}{\mat_1}_{\mu} = -32 e^2 g^2 \frac{2E(E-2q)}{[E-2w][E-2k]} . 
\end{equation}

\noindent Comparing the above three equations with the result from the loop 
calculation (Eq.\ref{lpphtndcy}, \ref{lp2phtndcy}) gives us the relations:

\begin{eqnarray}
Disc[{\Pi^{2}}^{\mu}_{\mu}](E-k-q-\kq) &=&  i \int \frac{ d^3 k d^3 q d^3 w }{ (2\pi)^{9} 2q 2k 2w }
(2 \pi )^4 \kd^4( \fp - \fk - \fq - \fw )
\nonumber \\
&\times & \Big\{ [ 1 - \nf(k) ][ 1 + n(q) ][ 1 - \nf(w) ] -  \nf(k) n(q) \nf(w) \Big\}
\nonumber \\
&\times & \Big[  { \mat_1^* }^{\mu} { \mat_1 }_{\mu} + { \mat_2^* }^{\mu} { \mat_2 }_{\mu} \Big],
\end{eqnarray}

\begin{eqnarray}
Disc[{\Pi^{1}}^{\mu}_{\mu}](E-k-q-\kq) &=&  i \int \frac{ d^3 k d^3 q d^3 w }{ (2\pi)^{9} 2q 2k 2w }
(2 \pi )^4 \kd^4( \fp - \fk - \fq - \fw )
\nonumber \\
&\times & \Big\{ [ 1 - \nf(k) ][ 1 + n(q) ][ 1 - \nf(w) ] -  \nf(k) n(q) \nf(w) \Big\}
\nonumber \\
&\times & \Big[ { \mat_2^* }^{\mu} { \mat_1 }_{\mu} + { \mat_1^* }^{\mu} { \mat_2 }_{\mu} \Big],
\end{eqnarray}

\noindent and hence we get the relation written down by Weldon \cite{wel83}

\begin{eqnarray}
Disc[{\Pi}^{\mu}_{\mu}](E-k-q-\kq) &=&  i \int \frac{ d^3 k d^3 q d^3 w }{ (2\pi)^{9} 2q 2k 2w }
(2 \pi )^4 \kd^4( \fp - \fk - \fq - \fw )
\nonumber \\
&\times & \Big\{ [ 1 - \nf(k) ][ 1 + n(q) ][ 1 - \nf(w) ] -  \nf(k) n(q) \nf(w) \Big\}
\nonumber \\
&\times & \Big[ { \mat^* }^{\mu} { \mat }_{\mu} \Big],
\end{eqnarray}

\noindent where $\mat = \mat^{\mu} \epsilon_{\mu}(p) = \mat_1 + \mat_2 $ 
is the full matrix element of heavy photon decay.


\subsection {Compton Scattering.}


The analysis for Compton scattering is slightly more tricky. 
Note that there are two sets of terms from 
Eqs.~(\ref{discAc},\ref{discDc}) and Eqs.~(\ref{discCc},\ref{discD2+3c},\ref{discD1c}) 
that may lead to Compton scattering. One appears with the 
delta function $\kd(E+k-q-\kq)$ and the other with the delta 
function $\kd(E+\kq-k-q)$. The delta functions can be turned 
into one another simply by replacing $ \vk \ra \vk  + \vq $, 
followed by $\vq \ra -\vq$. One notes on performing this operation 
that the rest of the integrand looks rather different. This happens 
as there are 4 topologically distinct diagrams that may fall under 
the category of Compton scattering ( it is well known that that for 
a given in state there are two diagrams that lead to Compton scattering; 
there are four here as we sum over the possibilities of the incoming 
fermion being a quark or antiquark). 
Let us consider the contribution from \po:

\begin{eqnarray} 
Disc[\Pi^{1}]( E + k - \qk - q ) &=& (2\pi i) 8 e^2 g^2  
\int \frac{ d^3 k d^3 q}{ (2\pi)^{6} q}  \nonumber \\
&\times & \Big \{ \nf(k) [1-\nf(\qk)] [ 1 + n(q) ] - [ 1 - \nf(k) ] \nf(\qk) n(q) \Big \}
\nonumber \\
& \times & \frac{s_3 s_4 [\hk_{-} \cdot \uqk_+] [\hk_{s_4} \cdot \uqk_{s_3}]}
{[ -k + q + s_3\qk ][ q + \qk + s_4k] } \kd(E+k-\qk-q). \label{cmptn1}
\end{eqnarray} 

For the contribution from \pt, recall that we have an overall 
factor of two on each of the results of 
Eqs.~(\ref{discCc},\ref{discD2+3c},\ref{discD1c}) coming 
from the overall symmetry factor of \pt being double that 
of \po. We take half of the contribution from the $\kd(E+k-\kq-q)$ 
term, and half from the $\kd(E+\kq-k-q)$ term, and in the second 
contribution change $ \vk \ra \vk  + \vq $, followed by 
$\vq \ra -\vq$. This gives the total contribution from \pt as

\begin{eqnarray} 
Disc[\Pi^{2}]( E + k - \qk - q ) &=& (2\pi i) 8 e^2 g^2  
\int \frac{ d^3 k d^3 q}{ (2\pi)^{6} q}  \nonumber \\
& \times & \Big \{ \nf(k) [1-\nf(\qk)] [ 1 + n(q) ] - [ 1 - \nf(k) ] \nf(\qk) n(q) \Big \}
\nonumber \\
& \times & \Bigg[ 
\frac{ \hk_{-} \cdot \uqk_+ }{[ k + q + \qk ]^2 } 
 +\frac{ \hk_{-} \cdot \uqk_+ }{[ k - q + \qk ]^2 }
\Bigg]
\kd(E+k-\qk-q) .\label{cmptn2}
\end{eqnarray}

\begin{figure}[htbp]
  \begin{center}
  \epsfxsize 120mm
  \epsfbox{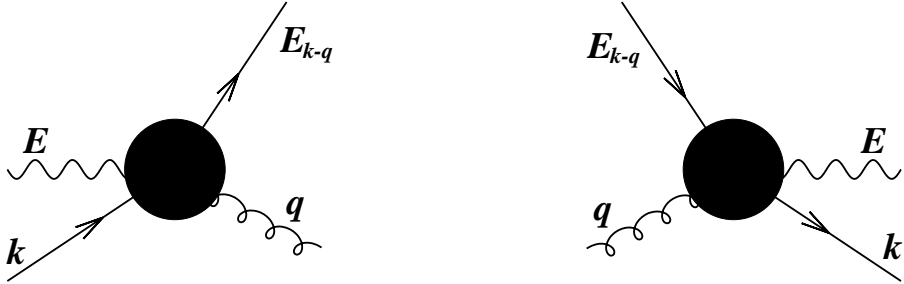}
\vspace{0.25cm}
    \caption{Quark Compton scattering.}
    \label{compton}
  \end{center}
\end{figure}

Notice that the combination of distribution functions appearing in 
the curly brackets are identical. The product of the three factors 
$\nf(k) [1-\nf(\qk)] [ 1 + n(q) ]$ has the interpretation of an 
incoming quark(or an antiquark) from the medium fusing with the 
photon coming in from outside the bath, resulting in an gluon and 
a quark(or antiquark) going into the medium. Subtracted from this 
is the product $[ 1 - \nf(k) ] \nf(\qk) n(q)$, which has the 
interpretation of an incoming quark(antiquark) from the medium 
fusing with an incoming gluon from the medium, resulting in a 
quark(antiquark) going back into the medium, and a virtual photon 
which leaves the medium (Fig.~\ref{compton}).  

To convert the above expressions into cross sections for Compton scattering, 
we again define the new four-vector $  \fw = (w,\vec{w})$, such that  $w = E + k - q$. 
This is generalized to 

\[
\fw = \fp + \fk - \fq.
\] 
as a result $\vec{w} = \vec{k} - \vec{q} $. Now, requiring that 
$\fw$ be on shell imposes the condition that 

\[
(E+k-q)^2 = k^2 + q^2 -2kq\cos{\h} 
\]

\[
\im E(E+2k-2q) = \fk \cdot \fq. 
\]

Using the above relations we many now rewrite the discontinuity obtained 
from \pt. In the numerators of the integrand we notice the factor 
$\wh{(k-q)}_+ \cdot \hk_-$, which may be written as  
$\hat{\fw}_+ \cdot \hat{\fk}_- = (E+2k)(E-2w)/2$. We introduce the standard denominators 
$2k,2w$ and factors of $2\pi$  and perform a similar set of operations as for photon decay 
to obtain the full result for Compton scattering as  
\begin{eqnarray}
Disc[\Pi](E+k-q-\kq) &=&  i \int \frac{ d^3 k d^3 q d^3 w }{ (2\pi)^{9} 2q 2k 2w }
(2 \pi )^4 \kd^4( \fp + \fk - \fq - \fw )
\nonumber \\
&\times & \Big\{ \nf(k)[ 1 + n(q) ][ 1 - \nf(w) ] -  [ 1 - \nf(k) ] n(q) \nf(w) \Big\}
\nonumber \\
&\times & 
32 e^2 g^2 \Bigg[ 
\frac{ E + 2k }{ E - 2w } + \frac{ E - 2w }{ E + 2k } 
+ 2 \frac{E(E-2q)}{[E-2w][E+2k]} 
\Bigg]. \label{lpcmptn}
\end{eqnarray}

\begin{figure}[htbp]
  \begin{center}
  \epsfxsize 93mm
  \epsfbox{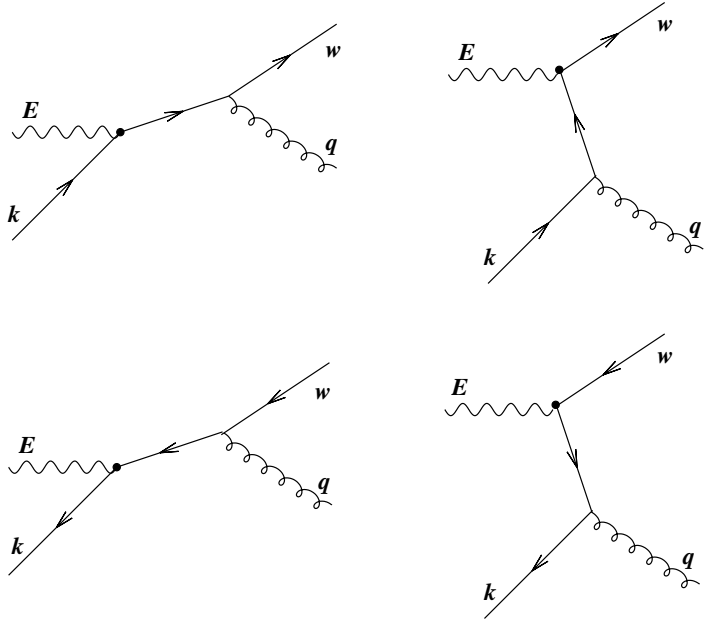}
\vspace{0.25cm}
    \caption{Compton scattering at first order in $\alpha$ and $\alpha_s$.}
    \label{compton1}
  \end{center}
\end{figure}

Recall that in \po there is another term, the coefficient of the 
delta function $\kd(E+\qk -k - q)$, that leads to Compton 
scattering. Also, in the Compton scattering contributions 
from \pt, we only used a half of both the terms. Following 
almost the same method as above, one can demonstrate that 
the form of the contribution from these terms is almost the 
same as above with $k$ and $w$ interchanged. In it, one may 
interchange $\vw \ra \vk$ to get the same contribution as 
Eq. (\ref{lpcmptn}); hence doubling the total contribution 
from Compton scattering.

Compton scattering by an incoming photon of a thermal medium of quarks and antiquarks, 
at first order in the E.M. and strong coupling constant can occur as a result of 
four processes as shown in Fig.~\ref{compton1}. The matrix element for 
the diagrams may be written as $\mat_{n}=\mat_{n}^{\mu} \epsilon _{\mu}(\fp)$, where

\[
\mat_{1}^{\mu} =  \bar{u}(\fw) ig \g^{\r} \e^*_{\r}(\fq) 
i \frac{\f\fp + \f\fk}{(\fp + \fk)^2} 
i e \g^{\mu} \e_{\mu} (\fp) u(\fk)
\]

\[
\mat_{2}^{\mu} =  \bar{u}(\fw) i e \g^{\mu} \e_{\mu}(\fp)
\frac{\f\fw - \f\fp}{(\fw - \fp)^2}
i g  \g^{\r} \e^*_{\r}(\fq) u(\fk).
\]

\noindent The amplitude for the third and fourth diagram can be obtained from 
the two amplitudes above simply changing $u \ra v$. Taking the 
products and summing over spins and colors 
(remember diagrams 1 and 2 interfere with each other, and so do 3 and 4), we get

\begin{equation}
{\mat^{*}_{1}}^{\mu}{\mat_1}_{\mu} = 32 e^2 g^2 \frac{E-2w}{E+2k},  
\end{equation}

\begin{equation}
{\mat^{*}_2}^{\mu}{\mat_2}_{\mu} = 32 e^2 g^2 \frac{E+2k}{E-2w} . 
\end{equation}

Once again, note that 
${\mat^{*}_{1}}^{\mu}{\mat_1}_{\mu},{\mat^{*}_2}^{\mu}{\mat_2}_{\mu}$ 
are negative. This is because $E-2w = q-k-w$ is always negative due 
to the triangle condition mentioned in the previous subsection. 
The cross term is 

\begin{equation}
{\mat^{*}_2}^{\mu}{\mat_1}_{\mu} = 32 e^2 g^2 \frac{2E(E-2q)}{[E-2w][E+2k]} . 
\end{equation}

Comparing the above three equations with the result from the 
loop calculation (Eq. (\ref{lpcmptn})) 
gives us the relation:

\begin{eqnarray}
Disc[{\Pi}^{\mu}_{\mu}](E+k-q-\kq) &=&  i \int \frac{ d^3 k d^3 q d^3 w }{ (2\pi)^{9} 2q 2k 2w }
(2 \pi )^4 \kd^4( \fp + \fk - \fq - \fw )
\nonumber \\
& \times & \Big\{ \nf(k)[ 1 + n(q) ][ 1 - \nf(w) ] - [ 1 - \nf(k) ] n(q) \nf(w) \Big\}
\nonumber \\
& \times & \Big[ {\mat^{*}_{1}}^{\mu}{\mat_1}_{\mu} + {\mat^{*}_2}^{\mu}{\mat_2}_{\mu} +
2 {\mat^{*}_2}^{\mu}{\mat_1}_{\mu} \Big].
\end{eqnarray}

Once again we note the interesting fact that in this gauge the mixed terms 
${ \mat_2^* }^{\mu} { \mat_1 }_{\mu} + { \mat_1^* }^{\mu} { \mat_2 }_{\mu}$ 
are always given by \po and the square terms 
${ \mat_1^* }^{\mu} { \mat_1 }_{\mu} + { \mat_2^* }^{\mu} { \mat_2 }_{\mu}$ 
are furnished by \pt. 


\subsection {Pair Creation}


\begin{figure}[htbp]
  \begin{center}
  \epsfxsize 100mm
  \epsfbox{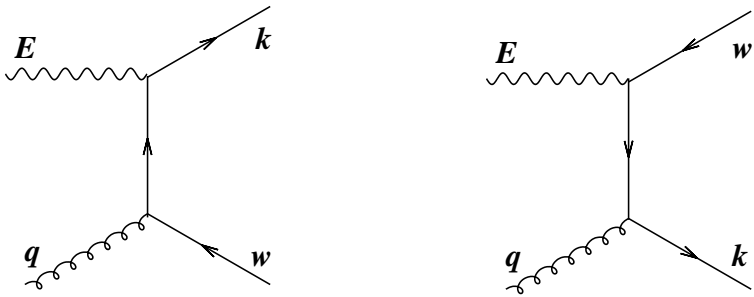}
\vspace{0.25cm}
    \caption{Pair creation at first order in $\alpha$ and $\alpha_s$.}
    \label{pair}
  \end{center}
\end{figure}

The analysis for pair creation(often referred to as photon-gluon fusion) 
is almost identical to the two previous sections. Its contribution is 
furnished by the only remaining delta functions in 
Eqs.~(\ref{discAc},\ref{discDc}) in the first self-energy, and 
Eqs.~(\ref{discCc},\ref{discD2+3c},\ref{discD1c}) in the second 
self-energy, \ie, $\kd(E+q-k-\kq)$. We simply state the results here: 
pair creation can occur through two types of processes, and has the 
discontinuity in the total self energy as

\begin{eqnarray}
Disc[\Pi^{2}](E+q-k-\kq) &=&  -i \int \frac{ d^3 k d^3 q d^3 w }{ (2\pi)^{9} 2q 2k 2w }
(2 \pi )^4 \kd^4( \fp - \fk + \fq - \fw )
\nonumber \\
&\times & \Big\{ [ 1 - \nf(k) ]n(q)[ 1 - \nf(w) ] -  [ 1 + n(q) ] \nf(k) \nf(w) \Big\}
\nonumber \\
&\times & 32 e^2 g^2 \Bigg[ \frac{ E - 2k }{ E - 2w } + \frac{ E - 2w }{ E - 2k }  + 
2 \frac{E(E+2q)}{[E-2w][E-2k]} \Bigg]. \label{lppair}
\end{eqnarray}


\section{PHYSICAL INTERPRETATION: LOOP-CONTAINING CUTS}

\begin{figure}[htbp]
  \begin{center}
  \epsfxsize 80mm
  \epsfbox{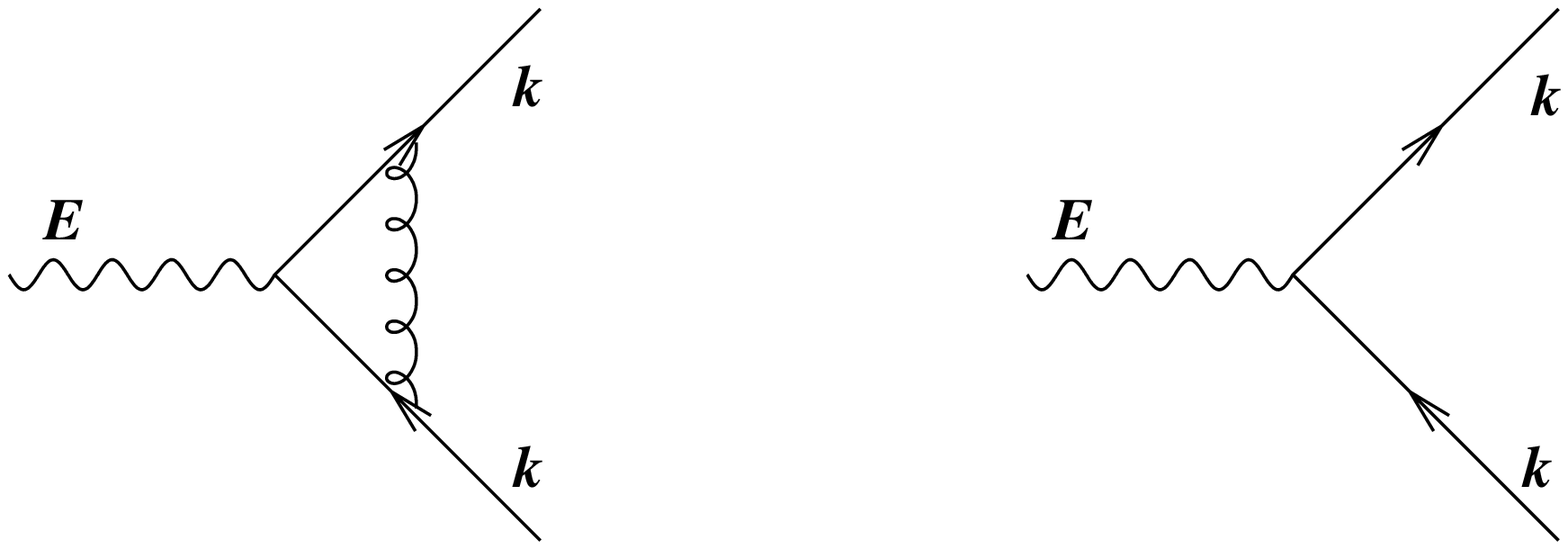}
\vspace{0.25cm}
    \caption{Photon decay at one loop corresponding to the cut $\kd(E-2\kq)$.}
    \label{loop1}
  \end{center}
\end{figure}
\begin{figure}[htbp]
  \begin{center}
  \epsfxsize 80mm
  \epsfbox{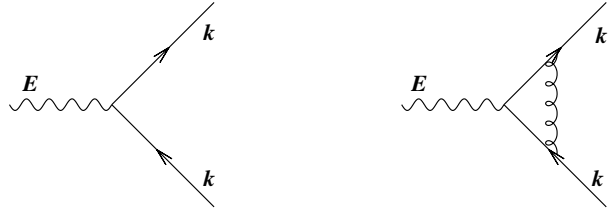}
\vspace{0.25cm}
    \caption{Photon decay at one loop corresponding to the cut $\kd(E-k)$.}
    \label{loop2}
  \end{center}
\end{figure}

We now analyze the various discontinuities of \po and \pt which contain 
loops. We start with the discontinuities of \po. These are given by 
Eqs.~(\ref{discAa},\ref{discAb},\ref{discDa}). We note that there are 
two terms with the delta function $\kd(E-2\kq)$ these correspond to the 
cut of Fig.~\ref{loop1}.
There is, also, one term with the cut $\kd(E-2k)$, this corresponds to 
the cut of Fig.~\ref{loop2}.

One may be satisfied with this interpretation of the cut diagrams and 
not proceed further.
A recent paper \cite{won01}, however,  has drawn attention to the fact that 
one can obtain a 
somewhat different interpretation of these diagrams, in terms of 
interference between 
simple tree like diagrams and diagrams containing particles 
called ``spectators''. 
Spectators are essentially on-shell particles from the heat bath 
that enter with the in-state  and leave with the out-state without 
having interacted with the the rest of the ``participants''.

We start by summing over the variable $s_1$ in Eq.~\ref{discAa}. This immediately gives
two terms, distinguished by the combination of distribution functions they carry:

\begin{eqnarray}
Disc[2\Pi_{\mu}^{\mu}(A)]_{a1} &=&  (-2 \pi i) 8 e^2 g^2  
\int \frac{ d^3 k d^3 q}{ (2\pi)^{6} }
\frac{  [\hk_+ \cdot \uqk_{s_2}][ \hk_{-} \cdot \uqk_{s_3} ] }
{  q ( E - (s_2-s_3)E_{q-k} )}  \nonumber \\
& \times & s_2 \frac{ [ 1 - 2\nf(k) ][ 1 - 2\nf(\qk) ]q }{ ( k + s_3\kq )^2 - q^2 } \kd(E-2k).
\label{discAa1}
\end{eqnarray}

and 

\begin{eqnarray}
Disc[2\Pi_{\mu}^{\mu}(A)]_{a2} &=&  (-2 \pi i) 8 e^2 g^2  
\int \frac{ d^3 k d^3 q}{ (2\pi)^{6} }
\frac{  [\hk_+ \cdot \uqk_{s_2}][ \hk_{-} \cdot \uqk_{s_3} ] } 
{  q ( E - (s_2-s_3)E_{q-k} )} \kd(E-2k) \nonumber \\
& \times & (-s_2) [ 1 - 2\nf(k) ][ 1 + 2n(q) ] \frac{ s_3k + \qk }{ ( k + s_3\kq )^2 - q^2 } .
\label{discAa2}
\end{eqnarray}

In Eq.\ref{discAa1}, if we replace $s_2 \ra -s_4$,$s_3 \ra -s$, followed by 
$\vk \ra \vq - \vk$ we get 

\begin{eqnarray}
Disc[2\Pi_{\mu}^{\mu}(A)]_{a1} &=&  (+2 \pi i) 8 e^2 g^2  
\int \frac{ d^3 k d^3 q}{ (2\pi)^{6} }
\frac{  [\uqk_- \cdot \hk_{s_4}][ \uqk_{+} \cdot \hk_{s} ] }
{  q ( E - (s-s_4)E_{q-k} )}  \nonumber \\
& \times & s_4 \frac{ [ 1 - 2\nf(k) ][ 1 - 2\nf(\qk) ]q }{ ( \qk - sk )^2 - q^2 } \kd(E-2\qk).
\label{discAa1ch}
\end{eqnarray}
 
The above is exactly the same as Eq.(\ref{discAb}). 
This is to be expected as the two cuts should 
in principle 
represent the same diagram up to a shift in momenta. 
We thus double this contribution and focus on it.
It represents photon decay into two quarks with quark emission and 
absorption from the final state
quarks. The other part from
Eq.~(\ref{discAa2}) along with Eq.~(\ref{discDa}) will represent photon 
decay with gluon emission and absorption off the external quarks. 


\subsection{ Photon decay with quark emission-absorption off vertex and final state. } 


We begin by summing over the remaining sign variables $s_2,s_3$ in 
$Disc[2\Pi_{\mu}^{\mu}(A)]_{a1}$ to get

\begin{eqnarray}
Disc[2\Pi_{\mu}^{\mu}]_4 &=&  2 \times (-2 \pi i) 8 e^2 g^2  
\int \frac{ d^3 k d^3 q}{ (2\pi)^{6} }
[ 1 - 2\nf(k) ][ 1 - 2\nf(\qk) ]\kd(E-2k)
\nonumber \\
& \times & \Bigg[ 
\frac{[\hk_+ \cdot \uqk_{+}][ \hk_{-} \cdot \uqk_{+} ] }
{E [ ( k + \kq )^2 - q^2 ]} 
- \frac{[\hk_+ \cdot \uqk_{-}][ \hk_{-} \cdot \uqk_{+} ] }
{(E+2\qk) [ ( k + \kq )^2 - q^2 ]}
\nonumber \\
& & + \frac{[\hk_+ \cdot \uqk_{+}][ \hk_{-} \cdot \uqk_{-} ] }
{(E-2\qk) [ ( k - \kq )^2 - q^2 ]}
- \frac{[\hk_+ \cdot \uqk_{-}][ \hk_{-} \cdot \uqk_{-} ] }
{E [ ( k - \kq )^2 - q^2 ]}
\Bigg].
\label{qrkspec}
\end{eqnarray}

As in the previous section, the distribution functions will be reorganized 
to allow for an interpretation in terms of thermal weights for particle 
emission and absorption.  
In the first two terms we define the new 
lightlike four-vector $\fw$ such that 
$\vw = \vq-\vk$. In the last two terms we 
define $\fw$ such that $\vw = -\vq + \vk$. 
This allows us to change the variable of 
integration as $d^3 q \ra d^3 w $, as $k$ is a constant as
far as the $q$ integration is concerned. 
We may also redefine the distribution functions as

\[
[ 1 - 2\nf(k) ][ 1 - 2\nf(w) ] = \Big[ [ 1 - \nf(k) ][ 1 - \nf(k) ] - \nf(k) \nf(k) \Big]  
\Big[ [ 1 -  \nf(w) ]  - [ \nf(w) ] \Big].
\]

The first set of factors in the larger square brackets has the usual interpretation
\cite{wel83} of the thermal factors that are associated with the probability of
particle emission into a heat bath or particle absorbtion from a heat bath. In this
case they carry the obvious meaning of: 
[emit fermion of energy k][emit fermion of energy k]  
$-$ [absorb fermion of energy k][absorb fermion of
energy k].
The reader will note that unlike the self-energy cuts 
considered in \cite{wel83} or those
of the previous section, the two cut diagrams that will result from this 
imaginary part of the self-energy will not be symmetric, in 
the sence that it will be the interference between a diagram with a 
loop and a simple tree diagram. The thermal factors discussed above 
will be the same for either diagram as they pertain to the quark and 
anti-quark that emanate from the decay of the photon (or those that 
combine to form the photon). Both amplitudes that result from this 
imaginary part contain this process and 
thus have identical thermal factors. 

\begin{figure}[htbp]
  \begin{center}
  \epsfxsize 100mm
  \epsfbox{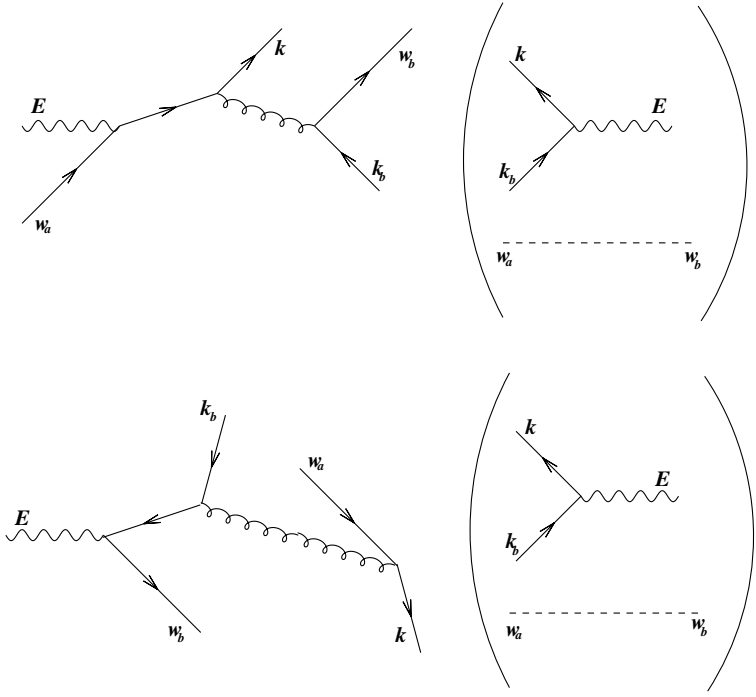}
\vspace{0.25cm}
    \caption{Interference between diagrams of different order in $\A_s$. The
diagrams on the left indicate $2 \ra 3$ reactions like 
$\g Q \ra q \bar{q} Q$ (where the $Q$ indicates that the incoming and outgoing
quarks are identical). The diagrams on the right indicate 
the complex conjugate
of Born term photon decay with a comoving quark spectator \ie, 
$(\g \ra q \bar{q})\otimes(Q \ra Q)$.} 
    \label{specfig1}
  \end{center}
\end{figure}

The second set of thermal factors has a new interpretation. These thermal factors
pertain to the particles in the remaining loop and 
thus are germane to only one of the two 
interfering amplitudes. 
We will demonstrate that these signal the difference of two amplitudes: 
that for the emission of a quark 
or anti-quark into the bath and its subsequent absorption from the bath, 
and vice-versa. Thus the second set of distribution functions is to be 
understood as: 
[emit fermion of four-momentum  w][absorb the same fermion of 
four-momentum  w]
$-$ [absorb fermion of four-momentum  w][emit the same fermion of 
four-momentum  w]. 

The process of emission of a fermion of four-momentum $\fw$ 
into a bath followed by its reabsorption is formally achieved 
by the action of creation and annihilation operators on the 
bath state $| \nf_{\fw} \rangle$ i.e,

\begin{equation}
a a^{\dag} | \nf_{\fw} \rangle = \Big( 1 - \nf_{\fw} \Big) | \nf_{\fw} \rangle.
\end{equation}
The reverse process, i.e., the absorption of a fermion from the bath and 
it subsequent re-emission into the bath is formally achieved by the action 
of annihilation and creation operators on the bath state i.e.,

\begin{equation}
a^{\dag} a | \nf_{\fw} \rangle = \Big( \nf_{\fw} \Big) | \nf_{\fw} \rangle.
\end{equation}

The discontinuity of the 
self-energy will represent the
amplitude of a particular process multiplied with the complex conjugate 
of another. 
In one of these
processes the above mentioned fermion will perform the emission and 
absorption procedure referred to
above. In the other amplitude, as we will show shortly, it will 
simply enter and leave without having
interacted with the rest of the particles. Due to this reason, 
it has been referred to previously (see 
\cite{won01,kap00}) as a spectator. 

We introduce the usual denominators of $2w,2k$ to get  

\begin{eqnarray}
Disc[\Pi_{\mu}^{\mu}]_4 &=&  2 \times (-2 \pi i) 8 e^2 g^2  
\int \frac{ d^3 k d^3 \w}{ (2\pi)^{6} kwkw }
\kd(E-2k)
\nonumber \\
&\times & \Big[ [ 1 - \nf(k) ][ 1 - \nf(k) ] - \nf(k) \nf(k) \Big] \nonumber \\
&\times  & \Big[ [ 1 -  \nf(w) ] - [ \nf(w) ] \Big]
\nonumber \\
&\times  & \Bigg[ 
\frac{[\fk_+ \cdot \fw_{+}][ \fk_{-} \cdot \fw_{+} ] }
{E [ ( k + w )^2 - q^2 ]} 
- \frac{[\fk_+ \cdot \fw_{-}][ \fk_{-} \cdot \fw_{+} ] }
{(E+2w) [ ( k + w )^2 - q^2 ]} 
\nonumber \\
& & + \frac{[\fk_+ \cdot \fw_{-}][ \fk_{-} \cdot \fw_{+} ] }
{(E-2w) [ ( k - w )^2 - q^2 ]}
- \frac{[\fk_+ \cdot \fw_{+}][ \fk_{-} \cdot \fw_{+} ] }
{E [ ( k - w )^2 - q^2 ]}
\Bigg].
\end{eqnarray}

We now introduce the new four-vector $\fk_b=(k,-\vk)$, and generalize the delta function to a 
four delta function. We then combine the first two terms and the last two terms to write

\begin{eqnarray}
Disc[\Pi_{\mu}^{\mu}]_4 &=&  i 8 e^2 g^2  
\int \frac{ d^3 k d^3 w d^3 k_b }{ (2\pi)^{9} 8 kwk_b }
16 (2\pi)^4 \kd^4( \fp + \fw - \fk - \fk_b - \fw )
\nonumber \\
&\times  & \Big[ [ 1 - \nf(k) ][ 1 - \nf(k_b) ] - \nf(k) \nf(k_b) \Big] \nonumber \\
&\times  & \Big[ [ 1 -  \nf(w) ] - [ \nf(w) ] \Big]
\nonumber \\
&\times  & \Bigg[ 
\frac{[\fk_b \cdot \fw ][ \fk_a \cdot ( \fp + \fw ) ] }
{[(E+w)^2 - w^2] [ ( k + w )^2 - q^2 ]} 
\nonumber \\
&+&  \frac{[\fk_a \cdot \fw ][ \fk_b \cdot ( \fw - \fp ) ] }
{ [ ( w - E )^2 - w^2 ][ ( k - w )^2 - q^2 ]}
\Bigg]. \label{spec1}
\end{eqnarray}
  
The above, has the interpretation of Fig.~\ref{specfig1}. This 
indicates the interference between 
two diagrams of different order in coupling constants.  
Let the matrix elements of the two tree-level 
diagrams with two propagators be denoted as $\mat_{1}=\mat_{1}^{\mu} \epsilon _{\mu}(\fp)$ 
and $\mat_{2}=\mat_{2}^{\mu} \epsilon _{\mu}(\fp)$. The matrix element of the term in brackets 
is simply denoted as $m^{\mu} \epsilon _{\mu}(\fp)$. Where the dotted line called the spectator is 
simply a product of Dirac delta functions over four momenta and Kronecker delta functions over 
the spins 
and colours of the incoming and outgoing fermions denoted by $w_a$ and $w_b$ (here, for brevity
 we indicate all the different quantum numbers, both continuous and discreet, of the incoming and
 outgoing particles by a single label).  

It is now simple to verify that the result obtained in Eq.~(\ref{spec1}) can be written as

\begin{eqnarray}
Disc[\Pi_{\mu}^{\mu}]_4 &=&  i 
\int \frac{ d^3 k d^3 w d^3 k_b }{ (2\pi)^{9} 8 kwk_b }
(2 \pi)^4 \kd^4( \fp + \fw - \fk - \fk_b - \fw )
\nonumber \\
&\times & \Big[ [ 1 - \nf(k) ][ 1 - \nf(k_b) ] - \nf(k) \nf(k_b) \Big] \nonumber \\
&\times & \Big[ \{ 1 -  \nf(w) \} - \nf(w)  \Big]
\nonumber \\
&\times & \Bigg[ 
 2{m^{\mu}}^*\mat_{1}^{\mu} + 2{m^{\mu}}^*\mat_{2}^{\mu}
\Bigg]. 
\end{eqnarray}

\noindent Where the Kronecker and Dirac delta functions over the 
fermions $w_a$ and $w_b$ have been used to set 
$w_a=w_b=w$. The factor of 2 preceeding the interference 
matrix elements is due to the fact that a similar 
process may be obtained by replacing an incomming quark 
spectator with an anti-quark spectator.


\subsection{Photon decay with gluon emission-absorption from final state quarks. } 

\begin{figure}[htbp]
  \begin{center}
  \epsfxsize 120mm
  \epsfbox{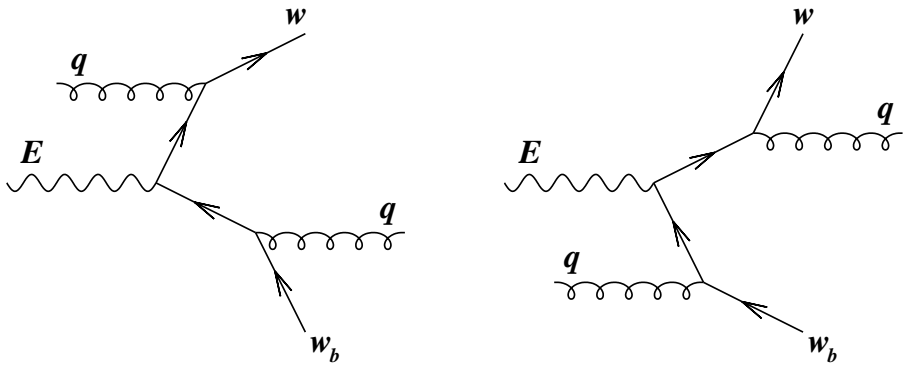}
\vspace{0.25cm}
    \caption{Photon decay with spectator gluon (the Born term with
spectator gluon is implied, see Fig.~\ref{specfig1}).}
    \label{specfig2}
  \end{center}
\end{figure}

This term receives contributions from  $Disc[2\Pi_{\mu}^{\mu}(A)]_{a2}$ and 
$Disc[2\Pi_{\mu}^{\mu}(D)]_{b}$. The fate of this discontinuity is 
essentially similar to the 
previous section and results once again in
the interference of tree level diagrams of different order. 
There are two sets of diagrams with two
propagators here as well, the difference being that the 
incoming, outgoing particle with the same set
of quantum numbers or in other words the spectator is a 
gluon. We once again introduce the on-shell
four-vector $\fw$, such that $\vw = \vk - \vq$. We use 
this to change the variable of
integration in $Disc[2\Pi_{\mu}^{\mu}(D)]_{b}$. 
$Disc[2\Pi_{\mu}^{\mu}(A)]_{a2}$ we relabel the dummy
variable $\vk \ra \vw$. Both discontinuities give essentially 
the same contribution thus the total
discontinuity from such processes (Fig.~\ref{specfig2}) is given as

\begin{eqnarray}
Disc[\Pi_{\mu}^{\mu}]_5 &=&  i 8 e^2 g^2  
\int \frac{ d^3 w d^3 q d^3 w_b }{ (2\pi)^{9} 8 kwk_b }
32 (2\pi)^4 \kd^4( \fp + \fq - \fw - \fw_b - \fq )
\nonumber \\
&\times & \Big[ [ 1 - \nf(w) ][ 1 - \nf(w_b) ] - \nf(w) \nf(w_b) \Big] \nonumber \\
&\times & \Big[ [ 1 +  n(q) ] + [ n(q) ] \Big]
\nonumber \\
&\times & \Bigg[ 
\frac{[\fw_b \cdot (\fq - \fw) ][ \fw_a \cdot ( \fq - \fw_b ) ] }
{[(q-w)^2 - k^2] [ ( q + w )^2 - k^2 ]} 
\Bigg]. \label{spec2}
\end{eqnarray}
  
Which is once again equal to 

\begin{eqnarray}
Disc[\Pi_{\mu}^{\mu}]_5 &=&  i\int \frac{ d^3 w d^3 q d^3 w_b }{ (2\pi)^{9} 8 kwk_b }
(2\pi)^4 \kd^4( \fp + \fq - \fw - \fw_b - \fq )
\nonumber \\
&\times & \Big[ [ 1 - \nf(w) ][ 1 - \nf(w_b) ] - \nf(w) \nf(w_b) \Big] \nonumber \\
&\times & \Big[ [ 1 +  n(q) ] + [ n(q) ] \Big]
\nonumber \\
&\times & \Bigg[ 
 {m^{\mu}}^*\mat_{1}^{\mu} + {m^{\mu}}^*\mat_{2}^{\mu}
\Bigg] .
\end{eqnarray}
 
Where $m$ represents the same process as in the previous 
subsection. The amplitudes $\mat_{1}$ 
and $\mat_{2}$ represent the processes of Fig.~\ref{specfig2}. 
The interpretation of the first set of
distribution functions is the same as before \ie, emission 
and absorption of two particles of energy
$k$. The second term has the interpretation of a gluon 
spectator exactly identical to that of the
quark spectator in the earlier subsection, but with Pauli factors 
replaced with Bose factors. Note that, unlike in the previous section,
 there is no factor of 2 preceeding the matrix elements as the 
spectators are gluons.


\subsection{Photon decay with quark and gluon emission-absorption off
 the same quark line } 


We, now, begin the analysis of the last loop-containing cut.
 This is essentially given by the discontinuities of 
Eqs.~(\ref{discA},\ref{discCa}). Combining these two 
discontinuities, and writing
$k=E-k$ in the denominators of the terms from 
$Disc[\Pi_{\mu}^{\mu}(A)]_a$ (note that we have to double this
contribution as it emanates from the second 
self-energy diagram which has a symmetry 
factor of 2 more than the first self-energy diagram), we
get  

\begin{eqnarray}
Disc[\Pi_{\mu}^{\mu}]_6 &=& (2 \pi i) 8 e^2 g^2 \int \frac{ dk d\h d\phi
\sin{\h} d^3 q}
{ (2\pi)^{6} q}  \kd(E-2k) [ 1 - 2\nf(k) ]
\nonumber \\
&\times & \Bigg\{ \frac { 2k \N \Sc  }{ [ E - s_2\kq - s_1q - k
]}
+  \frac { k^2 (\N \Sc )\p  }{ [ E - s_2\kq - s_1q -k ]} \nonumber \\
& & + \frac{ k^2 \N \Sc  [ 1 + s_2 \kq\p ]}{ [ E - s_2\kq -s_1q - k]^2 } 
\Bigg\}. \nonumber \\
&=& (2 \pi i) 8 e^2 g^2 \int \frac{ dk d\h d\phi \sin{\h} d^3 q}{ (2\pi)^{6} q}
\kd(E-2k) [ 1 - 2\nf(k) ]
\nonumber \\
&\times & \frac{d}{dk} \Bigg\{ \frac { k^2 \N \Sc  }{ [ E - s_2\kq - s_1q - k ] } \Bigg\} .\label{disc6}
\end{eqnarray} 

The above term does not readily admit a physical interpretation, however, 
the infrared limit will be evaluated with the above expression as the starting 
point as it is formally correct. 
To try and obtain a physical interpretation from the expression given above, 
an integration by parts is performed to obtain the discontinuity as

\begin{eqnarray}
Disc[\Pi_{\mu}^{\mu}]_6 &=& (2 \pi i) 8 e^2 g^2 \int \frac{ dk d\h d\phi \sin{\h} d^3 q}
{ (2\pi)^{6} q} \Big\{ -\frac{d\kd(E-2k)}{dk} \Big\} [ 1 - 2\nf(E/2) ] 
\nonumber \\
&\times & \frac { k^2 \N \Sc  }{ [ E - s_2\kq - s_1q - k ] }
\nonumber \\
&=&
(2 \pi i) 8 e^2 g^2 \int \frac{ d^3 k d^3 q}
{ (2\pi)^{6} q}   
\frac { \N \Sc  }{ [ k - s_2\kq - s_1q ] } \nonumber \\
&\times &  \Big\{ 2 \kd \p (E-2k) [ 1 - 2 \nf(E/2) ] \Big\}. \label{selfE}
\end{eqnarray} 

Where, we have used the property that 

\[
\frac{d\kd(E-2k)}{dk} = 2 \frac{d\kd(E-2k)}{d(2k)} =  - 2 \frac{d\kd(E-2k)}{d(E)}
= - 2\kd \p (E-2k). 
\]

Interestingly, as an aside, we note that one may still obtain a 
physical interpretation 
of the above term in terms of spectators with retarded propagators. 
To obtain this, we expand the factor 
$\frac { \N \Sc  }{ [ k - s_2\kq - s_1q ] }$ by summing over $s_1$ and $s_2$.
Here, as expected, we will obtain a part dependent on Bose distribution 
functions and a part dependent on Fermi distribution functions. We will 
illustrate the physical interpretation using the part containing the 
Bose distribution functions. We begin by writing the delta function in 
Eq. (\ref{selfE}) using the following regulator:

\begin{eqnarray}
\kd(x) = \lim_{\e \ra 0} \frac{ \e } { x^2 + \e^2 }.
\end{eqnarray}
 
In this regulation scheme, we obtain

\begin{eqnarray}
\kd \p (x) = -2 \bigg[ 
\frac{ \kd(x) }{ x + i\e }  + i \pi \kd^{2} (x) \Bigg] .
\end{eqnarray}
 
Substituting the above relation in Eq. (\ref{selfE}) we get 

\begin{eqnarray}
Disc[\Pi_{\mu}^{\mu}]_6 &=& (2 \pi i) 16 e^2 g^2 \int \frac{ d^3 kd^3 q}
{ (2\pi)^{6} q}  (-2) \kd(E-2k) [ 1 - 2\nf(E/2) ] 
\nonumber \\
& \times & \frac { \N \Sc  }{ [ E - s_2\kq - s_1q - k ] }
\Bigg\{ \frac{ 1 }{ ( E - 2k ) + i\e } + i \pi \kd(E-2k) \Bigg\}. \label{self2}
\end{eqnarray}

We now write unity in the form of an integral as

\begin{eqnarray}
1 &=& \int^{\infty}_{\infty} d \w^0 \frac{1}{2} \Big\{ 
\kd ( \w^0 + k ) + \kd ( \w^0 + k ) \Big\} \nonumber \\
&=& \int^{\infty}_{\infty} d \w^0 k \kd ( {\w^{0}}^{2} - k^{2}). \label{int1}
\end{eqnarray}

Substituting Eq. (\ref{int1}) in Eq. (\ref{self2}), we obtain 

\begin{eqnarray}
Disc[\Pi_{\mu}^{\mu}]_6 &=& (2 \pi i) 16 e^2 g^2 \int d \w^0 (E/2) 
\kd ( {\w^0}^2 - (E/2)^2 ) \int \frac{ d^3k d^3 q}
{ (2\pi)^{6} q}  (-2) \kd(E-2k) [ 1 - 2\nf(E/2) ] 
\nonumber \\
& \times & \frac { \N \Sc  }{ [ E - s_2\kq - s_1q - k ] }
\Bigg\{ \frac{ 1 }{ ( E - 2k ) + i\e } + 2 \pi i \nf (\w^0) \kd (E-2k) \nonumber \\ 
& & \mbox{} - 2 \pi i (\nf (\w^0) - 1/2 ) \kd(E-2k) \Bigg\}  \nonumber \\
&=& ( 2 \pi i) (-16 e^2 g^2) \int d \w^0  
\int \frac{ d^3k d^3 q}
{ (2\pi)^{6} q} \kd ( {\w^0}^2 - (E/2)^2 ) \kd(E-2k) [ 1 - 2\nf(E/2) ] 
\nonumber \\
& \times & \frac { E \N \Sc  }{ [ E - s_2\kq - s_1q - k ] }
\Bigg\{ \frac{ 1 }{ ( E - 2k ) + i\e } + 2 \pi i \nf (\w^0) \kd (E-2k) \nonumber \\ 
& & \mbox{} - 2 \pi i (\nf (\w^0) - \h(-\w^0) ) \kd(E-2k) \Bigg\}. 
\label{self3}
\end{eqnarray}

In the above we have simply added and subtracted the factor 
$2 \pi i \nf (\w^0) \kd (E-2k)$  inside the curly brackets and 
rewritten $1/2$ as $\h(-\w^0)$, as the rest of the integrand is
an even function of $\w^0$. 

We now introduce the three vector part of $\w$ as 

\begin{eqnarray}
\kd( {\w^0}^2 - (E/2)^2 ) \kd ( E - k - k) &=&
\int d^3 \w \kd( {\w^0}^2 - (E/2)^2 ) \kd ( E - k - k) \kd^3 (-\vk -\vec{\w}) \nonumber \\
&=&  \int d^3 \w \kd ( {\w^0}^2 - |\vec{\w}|^2 ) 
\kd ( E - k - |\vec{\w}| ) \kd^3 (-\vk -\vec{\w}).
\end{eqnarray}

As stated previously we now concentrate on the part of 
$ \N \Sc / [ k - s_2\kq - s_1q ]  $ which depends 
on the Bose distribution function. This gives 

\begin{eqnarray}
Disc[\Pi_{\mu}^{\mu}]_6 &=& i\int \frac{d^4 \w}{(2\pi)^4}  
\int \frac{ d^3k d^3 q}
{ (2\pi)^{6} 2q 2k} 2\pi \kd ( {\w^0}^2 - |\vec{\w}|^2 ) 
(2\pi)^4 \kd^4( \fp - \fk - \fw ) \nonumber \\
&\times & \Big [ \{ 1 - \nf(k^0) \} \{ 1 - \nf(\w^0) \} - \nf(k^0) \nf(\w^0)  \Big ] 
\Big [ \{ 1 + n(q) \}  + n(q)  \Big ]
\nonumber \\
&\times & (-64e^2 g^2) E^2 \Bigg[ 
\frac { k( E - k - q ) - \vk \x (\vk-\vq) }{ ( E - k - q )^2 - | \vk - \vq |^2 }
+ \frac { k( E - k + q ) - \vk \x (\vk-\vq) }{ ( E - k + q )^2 - | \vk - \vq |^2 } 
 \Bigg] \nonumber \\
&\times & \Big\{ \tilde{\Delta}_{R} (\w) \Big\} \nonumber \\ 
&=&i\int \frac{d^4 \w}{(2\pi)^4}  
\int \frac{ d^3k d^3 q}
{ (2\pi)^{6} 2q 2k} 2\pi \kd ( {\w^0}^2 - |\vec{\w}|^2 ) 
(2\pi)^4 \kd^4( \fp - \fk - \fw ) \nonumber \\
&\times & \Big [ \{ 1 - \nf(k^0) \} \{ 1 - \nf(\w^0) \} - \nf(k^0) \nf(\w^0)  \Big ] 
\Big [ \{ 1 + n(q) \}  + n(q)  \Big ]
\nonumber \\
&\times & \Big[ m^{\mu *}\mat_{1,\mu} +  m^{\mu *}\mat_{2,\mu} \Big].
\label{self4}
\end{eqnarray}

\begin{figure}[htbp]
  \begin{center}
  \epsfxsize 105mm
  \epsfbox{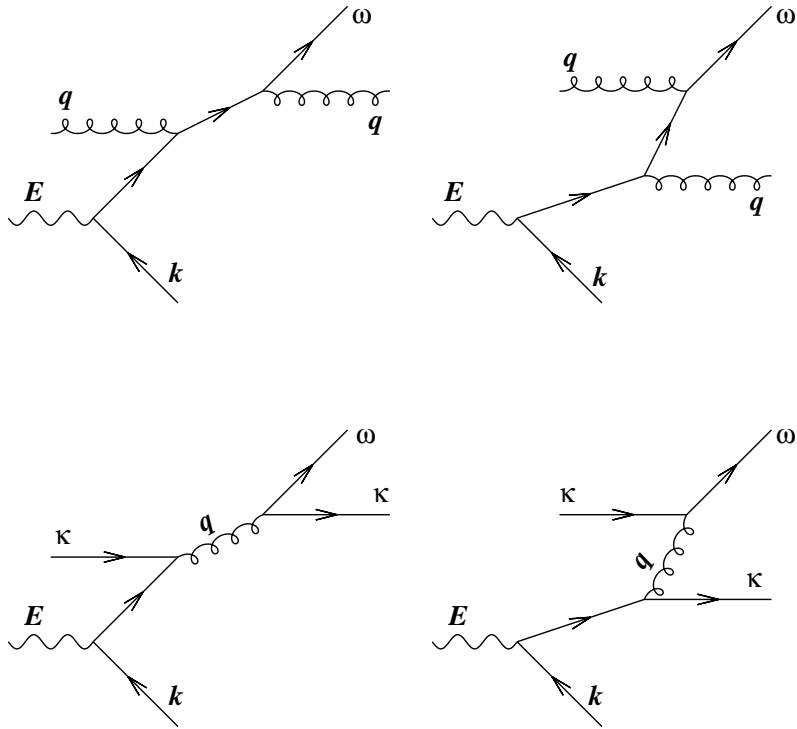}
\vspace{0.25cm}
    \caption{Photon decay into a $q \bar{q}$ pair. The quark then
emits or absorbs a quark or a gluon.}
    \label{specfig3}
  \end{center}
\end{figure}

Where $i\f\w\tilde{\Delta}_{R} (\w) $ is the retarded propagator. One may note that the 
integrand in the above equation is simply the interference matrix elements of the 
first ($\mat_{1} = \e^\mu(p) \mat_{1,\mu}$) and second diagram 
($\mat_{2} = \e^\mu(p) \mat_{2,\mu}$)of Fig.~\ref{specfig3} 
and the Born term ($m = \e_\mu(p) m^{\mu}$) with a gluon spectator.
A similar interpretation may be obtained for the the third and fourth 
diagrams of Fig.~\ref{specfig3} in terms quark spectators. 
However, as the above equation is not mathematically well defined; it 
will not be used in evaluating the infrared limit. Eq.~(\ref{disc6}) will
be used instead. 


\section{Infrared behaviour}
 

We now examine closely the infra-red and collinear singularity structure of the
terms enumerated in the two sections above. We will examine the infra-red
behaviour in the limit of heavy dilepton production from a plasma of massless
quarks \ie, $ E >> T $. There are essentially five terms:

$F$: Photon Gluon production denotes the reaction $q + \bar{q} \ra g + \g$ 

$C$: Compton like reaction  between a gluon and quark/anti-quark $g + q \ra \bar{q} + \g$

$D$: denotes the three body fusion to form the photon $g + q + \bar{q} \ra \g$

$V$: denotes photon formation from vertex corrected quark, antiquark.

$S$: denotes photon formation from self-energy corrected quark, anti-quark.

The full imaginary part of the two loop self-energy may be schematically written as

\begin{eqnarray}
2 \mbox{Im} \Pi^{\mu}_{\mu} 
&=& -\frac{8e^2g^2}{(2\pi)^3} \int dw \Big\{ n(w) 
[ F(w) + D_g(w) + V_g(w) + S_g(w) ] \nonumber
\\
&+& \tilde{n} (w) [ C(w) + D_q(w) + V_q(w) + S_q(w) ] \Big\}. \label{imgslfeng}
\end{eqnarray}

The first four terms represent the part of the terms mentioned above which are proportional to 
the gluon distribution function. The last four terms are those proportional to the
quark/antiquark distribution function. This is essentially the same notation as 
used by the authors of \cite{kap00b}. We now compute these contributions in turn.


\subsection{Self-Energy correction $S_g$ \& $S_q$}


The self-energy correction is essentially given by Eq. (\ref{disc6}), \ie,

\begin{eqnarray}
Disc[\Pi_{\mu}^{\mu}]_6 &=& (2 \pi i) 8 e^2 g^2 \int \frac{ dk d\h d\phi
\sin{\h} d^3 q}
{ (2\pi)^{6} q}  \kd(E-2k) [ 1 - 2\nf(k) ]
\nonumber \\
& &  \frac{d}{dk} \sum_{s_1,s_2}\frac { k^2 \N \Sc  }{ [ E - s_2\kq - s_1q - k ] }.
\end{eqnarray}
 
We concentrate, first, on the sum 
$\Sc = \sum_{s_1,s_2}\frac { k^2 \N \Sc  }{ [ E - s_2\kq - s_1q - k ] }$. This may be expanded
as 

\begin{eqnarray}
\Sc &=& \frac{k}{w} \Bigg[ 
\frac{[ 1/2 - \nf(w) + 1/2 + n(q) ] {\fw_+} \x {\fk_+} }{ E - k - w - q }
+ \frac{[ 1/2 - \nf(w) -(1/2 + n(q)) ] {\fw_+} \x {\fk_-} }{ E - k + w - q }
+ \mbox{} \nonumber \\
& & 
\!\!\!\!\!\!\!\!\!\!\!\!\!\!
\frac{[ -(1/2 - \nf(w)) + 1/2 + n(q) ] {\fw_+} \x {\fk_+} }{ E - k - w + q }
+ \frac{[ -(1/2 - \nf(w)) -(1/2 + n(q)) ] {\fw_+} \x {\fk_-} }{ E - k + w + q }
\Bigg]. \label{sg+sq}
\end{eqnarray}

We now concentrate on the terms proportional to $1/2 + n(q)$, \ie,

\begin{eqnarray}
\Sc_g = k \left[ \hf + n(q) \right] \Bigg[
\frac{ 2k( E - k - q ) - 2\vw \x \vk }{( E - k - q )^2 - w^2 }
+ \frac{ 2k( E - k + q ) - 2\vw \x \vk }{ ( E - k + q )^2 -w^2 }
\Bigg].
\end{eqnarray}

Introducing the variables $\A = E - 2k - 2q $, $\B = E -2k + 2q$ and $y = \cos \h$( where $\h$
is the angle between $\vk$ and $\vq$), we obtain

\begin{equation}
\Sc_g = k \left[ \hf + n(q) \right] \Bigg[
2 + \frac{ ( 2k - E ) \A }{ E\A + 2kq( 1 + y ) } + 
\frac{ ( 2k - E ) \B }{ E\B - 2kq( 1 - y ) } \Bigg].
\end{equation}

Dropping the $\hf$ ahead of the gluon distribution function we obtain the matter part of
$\Sc_g$. Using only this part we obtain (performing the unimportant angle integrations)

\begin{equation}
-\frac{8e^2g^2}{(2\pi)^3} \int  dq n(q) S_g(q) = 
\frac{8e^2g^2}{(2\pi)^3} \int  dk [1-2\nf(k)] \kd(k - E/2) \int dq q
\int dy \frac{d}{dk} \Sc_{ g, mat }.
\end{equation}

The limits of the $y$ integration are the locations for the onset of collinear singularities,
these are shielded by removing a small part of phase space $\e$ \ie, the $y$ integration is
performed within the limits $-1+\e \ra 1-\e$. The results will now depend on $\e$.  This gives
the result as

\begin{equation}
-\frac{8e^2g^2}{(2\pi)^3} \int  dq n(q) S_g(q) = 
\frac{-8e^2g^2}{(2\pi)^3} \int  dq n(q) \Bigg[ 
-4q - 4q \tbe \Bigg].
\end{equation}

In the above, the term $ 2\nf(E/2) $ has been dropped, as we are interested in the heavy dilepton limit
where $E>>T$ and as a result $ \nf(E/2) \ra 0 $. Thus, we get 

\begin{equation}
S_g(w) = -4w -4w \tbe. 
\end{equation}

We now concentrate on the terms, in Eq.~(\ref{sg+sq}), which are proportional to the factor 
$ 1/2 - \nf(w) $. Following a similar procedure as above we obtain 

\begin{equation}
-\frac{8e^2g^2}{(2\pi)^3} \int  dw \nf(w) S_q(w) = 
\frac{-8e^2g^2}{(2\pi)^3} \int  dw \nf(w) \Bigg[ 
-4w - 4w \tbe \Bigg]
\end{equation}

Thus, giving us the relation  

\begin{equation}
S_q(w) = -4w -4w \tbe .
\end{equation}

Note that in both the expressions for $S_q$ and $S_g$ there is a $\tbe$ term which blows up as $\e \ra 0$.
This is a collinear singularity. We shall allow $\e$ to vanish only when all the different contributions to
the dilepton rate have been added together.


\subsection{Vertex correction $V_g$ \& $V_q$}


We concentrate first on the term proportional to the fermionic 
frequency \ie, $V_q$. 
This vertex correction is essentially given by Eqs.~(\ref{discAa1}, 
\ref{discAa1ch}, \ref{qrkspec}).
The first two need to be doubled, as mentioned before in section 6. 
Extracting only the part proportional
to the fermionic distribution function $\nf(w)$, we obtain the $V_q$ integral
as

\begin{eqnarray}
-\frac{8e^2g^2}{(2\pi)^3} \int  dw \nf(w) V_q(w) &=& 2 \times  (-2\pi i) 8 e^2g^2 
\int \frac{ d^3k d^3w }{ (2\pi)^6 k^2 w^2}  \Big\{
[1-2\nf(k)][-2\nf(w)] \Big\} \nonumber \\
&\times & \Bigg[
\frac{ ( \fk_+ \x \fw_{s_2} )( \fk_- \x \fw_{s_3} )s_2 }
{ [ E - ( s_2 - s_3 )w ][ ( k + s_3w )^2 - q^2 ] } \Bigg]
\kd( E - 2k ). \label{qrkspec2}
\end{eqnarray}

Performing the sum on $s_2, s_3$ and setting $ y = \cos \h $(where $\h$ is the angle between $\vk$ and
$\vw$), we can perform one of the integrations with the help of one of delta function to get

\begin{eqnarray}
-\frac{8e^2g^2}{(2\pi)^3} \int  dw \nf(w) V_q(w) &=&  \frac{ 32 e^2 g^2 }{ ( 2\pi )^3 } \int dw w
(-\nf(w)) \int_{-1+\e}^{1-\e} dy \hf \Bigg[
\Bigg\{
\frac{ E }{ E + 2w } + \frac{ E }{ E - 2w } \Bigg\} \frac{ 1 + y }{ 1 - y } \nonumber \\
& & \mbox{} + \Bigg\{
\frac{ w }{ E + 2w } - \frac{ w }{ E - 2w } \Bigg\} ( 1 + y ) \Bigg].
\end{eqnarray} 

Note, once again, that the limits of the final angular integration $y$ signal the onset of collinear
singularities. These are, once again, regulated by removing the small part of phase space $\e$.
At this point we introduce the condition that the limit of interest is for dilepton mass much larger than
the temperature \ie, $E >> T$. The presence of the distribution function $\nf(w)$ on the energy $w$
severely restricts the contribution from regions where $w>>T$ to the integral. Thus, the dominant
contribution to the integral is from the regions where $w<<T$ or $w \sim T$. Hence, in the integral we may
make the approximation that $w<<E$ and expand the factors in the square brackets to linear power in $w/E$. 
This finally gives

\begin{eqnarray}
-\frac{8e^2g^2}{(2\pi)^3} \int  dw \nf(w) V_q(w) &=& -\frac{8e^2g^2}{(2\pi)^3} \int  dw \nf(w) \Bigg[ 
-8w + 8w \tbe \Bigg].
\end{eqnarray} 

Thus we obtain that

\begin{equation}
V_q(w) = -8w + 8w \tbe .
\end{equation}

Following almost a similar method as above we may obtain $V_g$ from 
Eq.~(\ref{discAa2}) (with an
overall factor of 2 as there is another cut which gives an identical contribution) as,

\begin{equation}
V_g(w) = 4w - \frac{2E^2}{w} \tbe .
\end{equation}

Once again, note that both expressions demonstrate a collinear divergence as $\e \ra 0$. The term $V_g$
also displays an infrared divergence as $w \ra 0$.


\subsection{Photon formation from quark, antiquark and gluon $D_g$ \& $D_q$}


The reverse reaction to this process represents heavy Photon ``decay'' into a $q \bar{q} \g$. Due to 
this reason, the process is denoted by the letter $D$ \cite{kap00b}. The full decay contribution is given
by Eq.~(\ref{lpphtndcy}) as,

\begin{eqnarray}
Disc[\Pi^{2}](E-k-q-\kq) &=&  -i \int \frac{ d^3 k d^3 q d^3 w }{ (2\pi)^{9} 2q 2k 2w }
(2 \pi )^4 \kd^4( \fp - \fk - \fq - \fw )
\nonumber \\
&\times & \Big\{ [ 1 - \nf(k) ][ 1 + n(q) ][ 1 - \nf(w) ] -  \nf(k) n(q) \nf(w) \Big\}
\nonumber \\
& \times& 32 e^2 g^2 \Bigg[ \frac{ E - 2k }{ E - 2w } + \frac{ E - 2w }{ E - 2k }  + 
2 \frac{E(E-2q)}{[E-2w][E-2k]} \Bigg]. 
\end{eqnarray}

In the above equation, note that if three of the delta functions are used to set 
$\vw = - \vk - \vq$, then
the remaining delta function imposes the condition that 

\[
E = k + q + \sqrt{ k^2 + q^2 + 2kq\cos\h }
\]

As mentioned before, we work in the limit $E>>T$, in this case the delta function can be satisfied by the
following regions of phase space:

a) $k \sim E$, $q \sim E$ and hence $w \sim E$; in this case all the distribution functions $n(q),\nf(k),
\nf(w) \ra 0$, and, thus, so do products of distribution functions.

b) $k \sim T<<E$, $q \sim E$ and hence $w \sim E$; 
in this case $\nf(k) \sim 1$. However $n(q), \nf(w) \ra 0$, and so do
products of distribution functions.

c) $w \sim T<<E$, $q \sim E$ and hence $k \sim E$; in this case 
$\nf(w) \sim 1$. However $n(q), \nf(k) \ra 0$, and so do
products of distribution functions. 

d) $q \sim T<<E$, $k \sim E$ and hence $w \sim E$; 
in this case $n(q) \sim 1$. However $\nf(w), \nf(k) \ra 0$, and so do
products of distribution functions.

Contributions from b) and c) will give us $D_q$, d) will give us $D_g$, while the contribution from a) is
negligible in comparison. We begin by calculating $D_q$ from the regions b) and c) of phase space. Here we
can ignore all combinations of distribution functions containing $n(q)$. As before, we also ignore the
vacuum term, concentrating only on the matter contribution. Noting the symmetry in the matrix element under
interchange of $k$ and $w$, we may change variables $w \ra k$ in the part of the integrand proportional to
the distribution function $\nf(w)$ to get

\begin{eqnarray}
-\frac{8e^2g^2}{(2\pi)^3} \int  dk \nf(k) D_q(k) &=&  \int \frac{ d^3 k d^3 q }{ (2\pi)^{5} 2q 2k 2w }
\kd( E - k - q - w )
\Big\{ 2\nf(k) \Big\}
\nonumber \\
& & 32 e^2 g^2 \Bigg[ \frac{ E - 2k }{ E - 2w } + \frac{ E - 2w }{ E - 2k }  + 
2 \frac{E(E-2q)}{[E-2w][E-2k]} \Bigg] .
\end{eqnarray} 
 
The argument of the delta function is the equation $g(q) = k + q + w(q) - E = 0$. 
The solution of this equation is at $q = q_{s}(k,E)$:

\begin{equation}
q_s = \hf \frac{ E( E - 2k ) }{ E - k( 1 - y ) }.
\end{equation}

The delta function can be written as

\[
\kd(g(q))  =  \frac{ \kd ( q - q_{s} ) }{ | g\p ( q_s ) | } .
\]

Substituting this back into the equation for $D_q$, we can do the $dq$ integration with the above mentioned
delta function. We can then perform the remaining angular integration by removing the small part of phase
space $\e$ to shield the collinear singularities. Now expanding up to linear order in $k$ as $k<<E$, we get

\begin{equation}
-\frac{8e^2g^2}{(2\pi)^3} \int  dk \nf(k) D_q(k) = -\frac{8e^2g^2}{(2\pi)^3} \int  dk \nf(k) \Bigg[
2k + \Big( - 2k - E 
\Big) \tbe \Bigg].
\end{equation}
  
Thus we get

\begin{equation}
D_q(w) = 2w + \Big( - 2w - E \Big) \tbe .
\end{equation}

We can now obtain $D_g$ by concentrating on region d) of phase space and 
ignoring all combinations of
distribution functions containing $\nf(k)$ or $\nf(w)$, this gives us,

\begin{equation}
D_g(w) = -2w + \Bigg( 2w - 2E  + \frac{E^2}{w} \Bigg) \tbe. 
\end{equation}


\subsection{Pair annihilation $F$ \& Compton scattering $C$}


The procedure to obtain these is almost exactly identical to the two terms
of the previous section. The total Compton scattering contribution can be obtained
from Eq.~(\ref{lpcmptn}) by doubling it as mentioned in the paragraph immediately 
following Eq.~(\ref{lpcmptn}). We may, once again, from phase space considerations 
show that the dominant contribution to Compton scattering occurs from a region where 
$k \sim T << E$(k is the incoming quark or antiquark energy). 
The leading term of Compton scattering is, thus, proportional
to the quark or antiquark distribution function.
From similar considerations the leading term of pair annihilation can be
demonstrated to be proportional to the outgoing 
gluon distribution function. Expanding them
up to linear order in the quark or gluon energy $w$, we get:

\begin{equation}
C(w) = 2w + \Big(
-2w + E \Big) \tbe.
\end{equation}

and

\begin{equation}
F(w) = -2w + \Bigg( 2w + 2E + \frac{E^2}{w} \Bigg) \tbe.
\end{equation}


\section{Results}


In the previous seven sections we evaluated the two different 
self-energies of the photon at two loops; then evaluated the various 
cuts of the self-energies which constituted its imaginary part; we 
then recombined the various cuts and reinterpreted them as physical 
processes; finally we evaluated these terms in the limit of 
heavy photon emission. In the last section we have concentrated solely 
on the thermal or matter part of these expressions. The vacuum part is
well known. All the expressions contain collinear singularities(as 
$\e \ra 0$), which 
for the moment have been shielded by removing the small part of phase 
space($\e$) where these singularities occur. Some of the expressions also 
display infrared singularities as $w \ra 0$. Hence, the final 
integrations over $w$ are yet to be performed. In the following we will
combine all these terms and perform this integration.

We now re-substitute the terms $F$, $C$, $D$, $V$, and $S$ back 
in Eq. (\ref{imgslfeng}) to get the coefficients of the bosonic and fermionic
distribution functions as

\begin{equation}
n(w)\Big( F + D_g + V_g + S_g \Big) = n(w)(-4w).
\end{equation}

\begin{equation}
\nf(w) \Big(C + D_q + V_q + S_q \Big) = \nf(w)(-8w).
\end{equation}

Thus, we find that when all the cuts are summed, the collinear and 
infrared singularities cancel. This is in contradistinction with the results of 
Refs. \cite{kap00,kap00b}, where the infrared singularities 
cancel but the collinear singularities persist. With these, we get the 
full imaginary part of the self-energy as

\begin{eqnarray}
\mbox{Im} \Pi^{\mu}_{\mu_{\mbox{2loop,thermal}}} &=& -\frac{4 e^2g^2}{(2\pi)^3} \Bigg[ 
- \frac{ 4 \pi^2 T^2 }{3} \Bigg] \nonumber \\
&=& \frac{ 8 e^2 \A_s T^2}{ 3 }.
\end{eqnarray}

We may also derive the Born term and quote the two-loop vacuum 
contribution(from \cite{fie89}) as   

\begin{equation}
\mbox{Im} \Pi^{\mu}_{\mu_{\mbox{1loop}}} + \mbox{Im} \Pi^{\mu}_{\mu_{\mbox{2loop,vacuum}}} = 
\frac{-3e^2}{2 \pi} E^2 \Bigg( 1 +
\frac{\A_s}{\pi} \Bigg).
\end{equation}


\section{Discussions and conclusions}


In this paper, we have calculated the imaginary 
part of the two-loop heavy boson retarded self-energy in the 
imaginary time formalism.  We also elucidate the analytic
structure of the self-energy by recombining and reinterpreting 
various cuts of the self-energy as physical processes which
occur in the the medium. Cuts with loops have been interpreted
as interference terms between $O(\A)$ tree scattering amplitudes    
and the Born term with spectators. At each stage the results from
the self-energy cuts was matched by re-deriving the amplitudes 
of the tree level diagrams. This constitutes an important
check of each of the contributions from the self-energy. 

Each of the
contributions contain infrared and collinear 
singularities. In the interest of simplicity, we analyzed this 
singular behaviour in the region where the dilepton mass is 
far greater than the temperature. This allows us to neglect a 
series of terms which appear sub-dominant. In each case we 
retained terms only up to order $T^2/E^2$. 
One might argue that this represents
a considerable approximation of the result. However the 
resulting simplification allows us to analyze far simpler 
and analytically integrable expressions. We would 
point out that this was precisely the approximation
used in \cite{kap00b,kap00} where a remnant collinear 
divergence was deduced at $O(T^2/E^2)$. When all the 
contributions were summed, all infrared 
and collinear divergences cancelled; leaving a finite result of $O(T^2/E^2)$. 
In this sense our results differ slightly from those of Ref.\cite{alt89} who find a 
remnant $O(T^4/E^4)$ result. The possible reasons for this discrepancy are 
many. For example, the authors of Ref.\cite{alt89} use a complicated finite 
temperature remormalization prescription, ours is the same prescription as 
at zero temperature; they apply finite self-energy corrections on the 
outer legs of all their processes and we do not. However, both calculations 
(as well as those of Ref. \cite{bai88}) yield the consistent result that 
all collinear and infrared divergences cancel in the final rate expression. 
This is consistent with the KLN theorem \cite{kin62,lee64}, even though 
a formal proof of the theorem at finite temperature is still 
elusive. We leave this and other extensions for future investigations.


\section{ACKNOWLEDGMENT}


The authors wish to thank Y. Aghababaie, P. Aurenche, A. Bourque, 
S. Das Gupta, S. Jeon, J. I. Kapusta, C. S. Lam, G. D. Mahlon and S. M. H. Wong for 
helpful discussions. A.M. acknowledges the generous support provided to 
him by McGill University through the Hydro-Quebec
fellowship.
This work was supported in part by the Natural
Sciences and Engineering Research Council of Canada and by { \it le fonds pour la
Formation de Chercheurs et l'aide \`{a} la Recherche du Qu\'{e}bec. }


\section{APPENDIX A}

{\centerline {\bf Notation}}


Our notation is categorized by the explicit presence of an apparent Minkowski time 
$x^{0} = -i \tau$ and a momentum $q^{0} = i2n \pi T$ or $ i (2n+1) \pi T$ for bosons or fermions
respectively. Our metric is 
$(1,-1,-1,-1)$.  For the case
of zero chemical potential our bosonic propagators have the same appearance as at zero
temperature, \ie, 

\begin{equation}
i\Delta(q) = \frac{i}{(q^{0})^{2} - |q|^2 }.
\end{equation}     

\noindent
The Feynman rules are also the same as at zero temperature, with the understanding that
we replace the zeroth component of the momentum by  $i (2n+1) \pi T$ for a fermion and
by an even frequency in the case of a boson. One may, in the case of zero chemical
potential, relate this to the familiar case of reference \cite{pis88} by noting that

\begin{equation}
\Delta(q) = \frac{1}{(q^{0})^{2} - |q|^2 } = \frac{-1}{(\omega_{n})^{2} + |q|^2 } 
= - \Delta_{E}(\omega_{n},q),
\end{equation}

\noindent
where $\Delta_{E}(\omega_{n},q)$ is the familiar Euclidean propagator presented in
the literature (\cite{pis88},\cite{kap89}). One may immediately surmise the form of
the non-covariant propagator $\Delta(|\vec{q}|,x^{0})$, the Fourier 
transform of which is the covariant propagator. 
 
\begin{eqnarray}
\Delta(q,q^{0}) &=& - \int_{0}^{\beta} d\tau 
e^{-i\omega_{n}\tau} \Delta_{E}(|\vec{q}|,\tau) \nonumber \\
&=& -i \int_{0}^{-i\beta} d x^{0} e^{iq^{0}x^{0}} \Delta_{E}(|\vec{q}|,\tau) 
\nonumber \\
&=& -i \int_{0}^{-i\beta} d x^{0} e^{iq^{0}x^{0}} \Delta(|\vec{q}|,x^{0}).
\end{eqnarray}

\noindent
The full fermionic propagators are  

\begin{equation}
S(q,q^{0}) = ( \gamma^{\mu} q_{\mu} ) (-i) \int_{0}^{-i\beta} d x^{0} 
e^{ i q^{0} x^{0} }
\Delta_{\mu} (|\vec{q}|,x^{0}) ,
\end{equation}

\noindent where,

\begin{equation}
\Delta_{\mu} (|\vec{q}|,x^{0}) = \frac{1}{2E_{q}} \sum_{s} f_{s} ( E_{q} )
e^{-isx^{0}( E_{q} ) }.
\end{equation}


\section{APPENDIX B}

{\centerline {\bf Discontinuity across a second order pole.} }


Imagine we have a function of a complex variable $F(z)$, and it is given to be in the form

\begin{equation}
F(z) = \int dx \frac{ f(z,x) }{ z - x } + \frac{ g(z,x) }{ ( z - x )^2 },
\end{equation}

where $x$ is a real variable, integrated on the real axis. Most of the discontinuities that
we evaluate can be cast in this general form. This can be rewritten as 

\begin{equation}
-F(z) = \int dx \frac{ f(z,x) }{ x - z } - \frac{ g(z,x) }{ ( x - z )^2 }, 
\label{general}
\end{equation}

The functions $f(z,x)$ and $g(z,x)$ are analytic in $x$ and hence admit a Taylor
expansion. 

\begin{equation}
f(z,x) = f(z,x=z) + \frac{df}{dx}(z,x=z)[ x - z ] + \frac{1}{2}
\frac{d^2f}{dx^2}(z,x=z)[ x - z ]^2 + ... \label{taylor}
\end{equation}

Substituting Eq.(\ref{taylor}) in Eq.(\ref{general}) we get

\begin{eqnarray}
-F(z) &=& \int dx \frac{ f(z,x=z) }{ x - z } + \frac{df}{dx}(z,x=z) + 
\frac{d^2f}{dx^2}(z,x=z)[ x - z ] + ... \nonumber \\
&-& \frac{ g(z,x=z) }{ ( x - z )^2 } -  \frac{dg}{dx}(z,x=z)\frac{1}{ x - z } 
-  \frac{1}{2} \frac{d^2g}{dx^2}(z,x=z)\frac{1}{ (x - z)^2 } - ...  
\label{gentylr}
\end{eqnarray}

Recalling that only the pure first order poles develop a discontinuity or imaginary
part at the pole we get the imaginary part of Eq.(\ref{gentylr}) as 

\begin{equation}
Disc[-F(z)] = \int dx 2\pi i \kd(x-z) \Bigg[ f(z,x) - \frac{dg}{dx}(z,x) \Bigg]. 
\end{equation}

\end{document}